\documentclass[natbib]{svjour3}                     
\smartqed  
\usepackage{wrapfig}
\usepackage[dvips]{graphics,graphicx}
\usepackage{amssymb}
\usepackage{mathptmx}      
\usepackage{color}

\def\msun{M_\odot}

\def\ga{\mathrel{\mathpalette\fun >}}
\def\simpropto{\lower.2ex\hbox{$\; \buildrel \sim    \over \propto \;$}}

\def\simpropto{\lower.2ex\hbox{$\; \buildrel \sim \over \propto \;$}}

\def\fun#1#2{\lower0.837ex\vbox{\baselineskip0ex\lineskip0.209ex
  \ialign{$\mathsurround=0ex#1\hfil##\hfil$\crcr#2\crcr\sim\crcr}}}

\newcommand{\aap}{{Astron. Astrophys.}}
\newcommand{\araa}{{Ann. Rev. Astron. Astrophys.}}
\newcommand{\apj}{{Astrophys. J.}}
\newcommand{\aj}{{Astron. J.}}
\newcommand{\apjl}{{Astrophys. J. Lett.}}

\newcommand{\prd}{{Phys. Rev. D}}
\newcommand{\pre}{{Phys. Rev. E}}

\newcommand{\nat}{{Nature}}
\newcommand{\jcap}{{J. Comp. Astrop. Phys.}}
\newcommand{\mnras}{{Monthly Not. Royal Astr. Soc.}}
\begin{document}

\title{Particle acceleration in relativistic outflows}

\titlerunning{Relativistic outflows}        

\author{Andrei Bykov \and Neil Gehrels \and Henric Krawczynski \and  Martin Lemoine \and Guy Pelletier \and Martin Pohl}  

\authorrunning{Bykov et al.} 

\institute{A.M.Bykov \at Ioffe Institute of Physics and Technology,
194021, St.Petersburg, Russia \email{byk@astro.ioffe.ru}\\
St.Petersburg State Politechnical University
 \and
 N. Gehrels \at
             NASA/Goddard Space Flight Center,
              Tel.: +301-286-6546,
              \email{neil.gehrels@nasa.gov}
\and
      H. Krawczynski \at
      Department of Physics, Washington University, St. Louis, MO 63130, USA
          \email{krawcz@wuphys.wustl.edu}
\and
           M. Lemoine \at
              Institut d'Astrophysique de Paris, CNRS - UPMC, 98bis
              boulevard Arago, 75014 Paris, France,
              \email{lemoine@iap.fr}
\and
      G. Pelletier \at
              Institut de Plan\'etologie et d'Astrophysique de Grenoble, France \\
              Tel.: +33-476-514570\\
              Fax: +33-476-448821\\
              \email{Guy.Pelletier@obs.ujf-grenoble.fr}           
           \and
M. Pohl \at
Institut f\"ur Physik und Astronomie,
Universit\"at Potsdam, 14476 Potsdam, Germany,
\email{marpohl@uni-potsdam.de} \\
DESY, 15738 Zeuthen, Germany, \email{martin.pohl@desy.de}
}

\maketitle

\begin{abstract}
  In this review we confront the current theoretical understanding of
  particle acceleration at relativistic outflows with recent
  observational results on various source classes thought to involve
  such outflows, e.g. gamma-ray bursts, active galactic nuclei, and
  pulsar wind nebulae. We highlight the possible contributions of
  these sources to ultra-high-energy cosmic rays.
\end{abstract}

\section{Introduction}
High-energy astrophysical phenomena stem from the generation of
powerful flows emanating from supernova explosions, gamma-ray bursts
(GRB), from ejections in the environment of black holes or neutron
stars that lead to the formation of very strong shocks, at which
particle acceleration takes place. The new developments in these
issues, especially for relativistic shocks, are based on the
interdependence between the shock structure, the generation of
supra-thermal particles and the generation of turbulence. It is
thought, and numerical simulations support that view, that the
penetration of supra-thermal particles in the shock precursor
generates magnetic turbulence which in turn provides the scattering
process needed for particle acceleration through the Fermi
process. This successful development, first elaborated for supernova
remnants (SNR), inspired similar investigations for the termination
shock of GRBs. However, in ultra-relativistic shocks, difficulties
arise with the transverse magnetic field that places a limitation to
particle penetration upstream and that drags particles in the
downstream flow and makes shock recrossing difficult. It turns out
that only sufficiently fast micro-turbulence can make the Fermi
process operative, as demonstrated by recent numerical
simulations. Following a review of the main observational results on
GRBs, active galactic nuclei (AGN), and pulsar wind nebulae (PWN),
these points are briefly discussed and astrophysical consequences are
drawn. We describe the role relativistic shocks inside relativistic
flows, e.g. the internal shocks of the prompt-emission stage of GRBs,
may play in the generation of ultra-high-energy cosmic rays
(UHECR). Noting that the energy required for supplying sub-GZK UHECR
is huge compared with the available energy budget, we also discuss
other sources, such as AGN and young pulsars, that may contribute to
the flux at ultra-high energies, all the more so if the composition is
enriched in heavy nuclei, as suggested by recent experimental results.

\section{Gamma Ray Bursts}
\label{grb}
The first GRB was observed by one of the {\it Vela}
satellites monitoring for the Nuclear Test
Ban Treaty in 1967, but the (unexpected)
astronomical results were not declassified and
published for another six years \citep{1973ApJ...182L..85K}. For many years
the nature of GRBs was unknown, since the distance scale
was completely unknown. Beginning about 20 years ago, the
cosmological spatial distribution of GRBs was strongly hinted
at due to the very isotropic distribution on the sky of GRBs
localized by {\it CGRO}.  The wealth of detailed information
garnered in the last seven years by {\it Swift}
\citep{2004ApJ...611.1005G} has taken the study of GRBs to the next level,
and indeed the current situation is in some sense more
confusing than our  naive pre-{\it Swift} picture
\citep[e.g.][]{2009ARA&A..47..567G}.

In this subsection we
    (i) review the basic properties of the two main
   types of GRBs, long and short,
   (ii) look at long GRBs in more detail,
  (iii) review the brief history of short GRBs and the difficulties
  entailed in their study,
   (iv) provide an overview of the acceleration processes  for GRB
   jets,
and (v) conclude with a recent
     results on high energy emission observed by {\it Fermi}.

 \subsection{GRB Properties}
\label{sec:grb1}

\begin{figure}[h!]
\begin{center}
\includegraphics[height=100truemm]{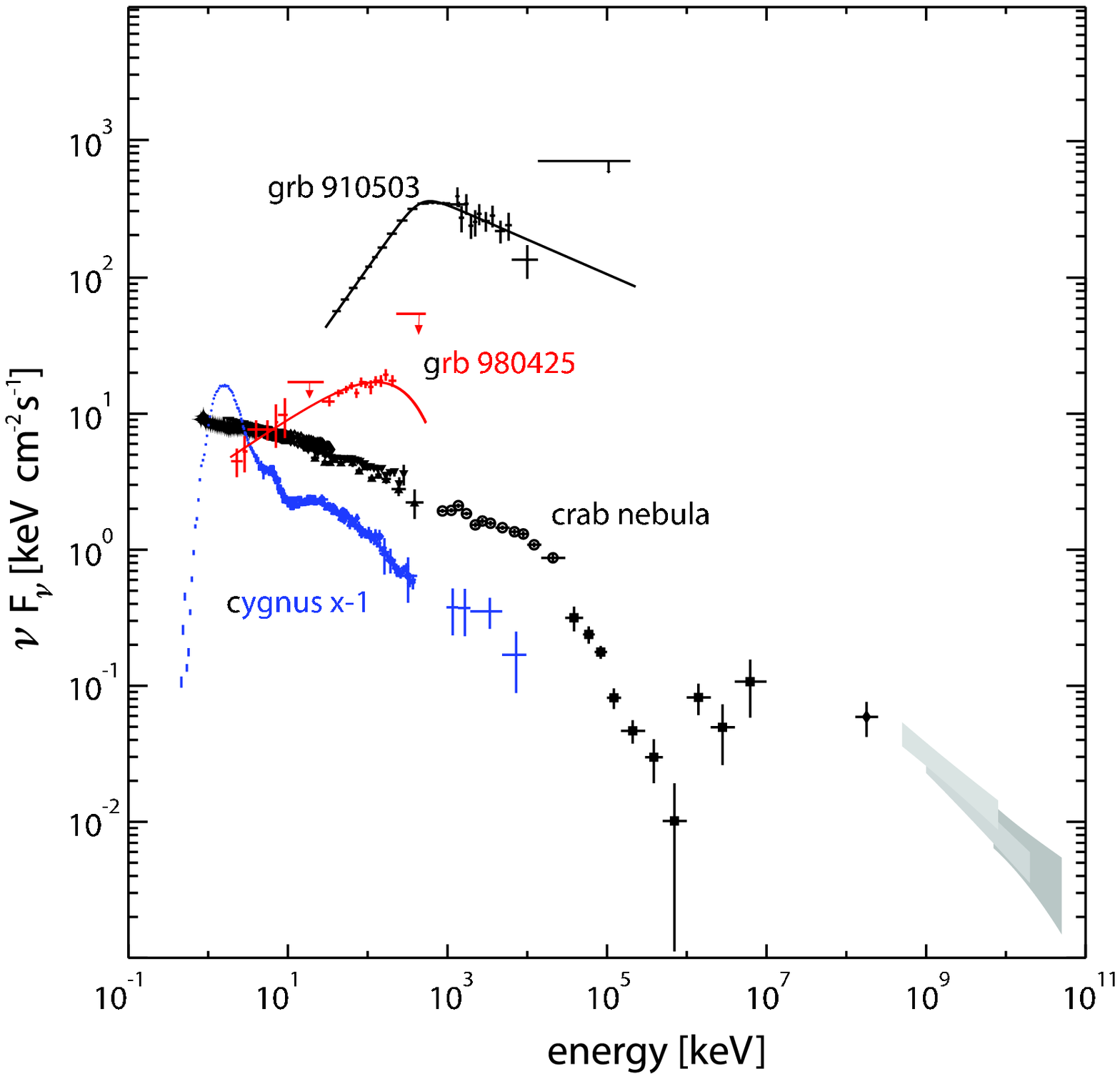}
\vskip -0.05 truein
\end{center}
 \caption{Representative broad-band $\nu F_{\nu}$ spectra
\citep{2009ARA&A..47..567G} of a lGRB (910503) \citep{2007ApJ...654..385K}
and a sGRB (980425) \citep{2008ApJ...677.1168K} along with the Crab pulsar nebula
\citep{2001A&A...378..918K} and Cyg X-1 \citep{2002ApJ...572..984M}.}
\label{fig:grb:spec}
\end{figure}

GRBs come in two kinds, long and short, where the dividing
line between the two is $\sim$2~s \citep{1993ApJ...413L.101K}.
Long GRBs (lGRBs) are thought to be due to
the collapse of a massive star, while short GRBs (sGRBs)
are inferred to be neutron star - neutron star (NS-NS) mergers.
A further division
can be made spectrally according to their hardness ratio
(i.e., ratio of high to low energies).
The redshift range is from
about 0.2 to 2 for sGRBs,
with a mean of about 0.4.
For lGRBs the range is between about 0.009 and 8.2, with a mean of about 2.3.
The typical energy release is $\sim$10$^{49}-10^{50}$ erg
for sGRBs and  $\sim$10$^{50}-10^{51}$ erg for lGRBs.
These ranges are based on observed isotropic-equivalent
energies of  $\sim$10$^{51}$ erg for sGRBs and $\sim$10$^{53}$ erg for lGRBs,
and estimates for jet beaming for each class,
$\theta_{\rm j}\sim5^{\circ}$ for lGRBs and
$\theta_{\rm j}\sim15^{\circ}$ for sGRBs
\citep{2006ApJ...653..468B,2006ApJ...653..462G}.
Beaming angles for sGRBs are still highly uncertain. The corresponding beaming factors
 $f_b = 1-\cos\theta_{\rm j}\simeq \theta_{\rm j}^2/2$
are  roughly $1/300$ for lGRBs and $1/30$ for sGRBs.
The sGRBs have weaker X-ray afterglows,
a mean value of
 $\sim$7$\times 10^{-10}$ erg cm$^{-2}$ s$^{-1}$  versus
 $\sim$3$\times 10^{-9}$  erg cm$^{-2}$ s$^{-1}$ for lGRBs.
Figure~\ref{fig:grb:spec} shows spectra for several
representative GRBs, and two other
high-energy sources, the Crab nebula and Cyg X-1.

\subsection{LGRBs}

The \emph{BeppoSAX}
mission made the critical discovery of X-ray afterglows of long
 bursts \citep{1997Natur.387..783C}.  With the accompanying discoveries by
ground-based telescopes of optical \citep{1997Natur.386..686V} and radio
\citep{1997Natur.389..261F} afterglows, lGRBs were found to emanate from
star forming regions in host galaxies at typical distance
of $z\simeq 1-2$. \emph{BeppoSAX} and the following \emph{HETE-2}
mission also found evidence of associations of GRBs with Type Ic
SNe.  This supported the growing evidence that lGRBs
are caused by ``collapsars'' where the central core of a massive
star collapses to a black hole \citep{1999ApJ...524..262M}.

LGRBs are incredibly bright.
A typical galaxy at a redshift of only $z=3$ is fainter than $m\simeq27$.
Multiwavelength observations of the current record holder, GRB 090432 (at $z\simeq8$),
are providing information about the universe at a
time when it was
only about 4\% of its current age, and shed light on the process of reionization
in the early universe \citep{2009Natur.461.1254T,2009Natur.461.1258S}.
The highest redshift GRBs are seen to have
high luminosity, resulting in fluxes
well above the detection threshold.
Such bursts are also strong at other wavelengths.
Table~\ref{relout:ta1}
presents optical data for the
highest redshift GRBs observed to date,
where the look-back time  $t_{\rm LB}$(Gyr)
is given in column 2.

\begin{center}
\centering
\begin{table}[h!]
\centering
\small
\centering
\caption{High $z$ GRBs.}
\begin{tabular}{@{}cccccc@{}}
\noalign{\smallskip}\hline\noalign{\smallskip}
 $z$     &     $t_{\rm LB}$(Gyr)  & GRB &  Optical Brightness &   &  \\
\noalign{\smallskip}\hline\noalign{\smallskip}
8.3 & 13.0 & 090423 &  $K=20$  & @ & 20 min  \\
6.7 & 12.8 & 080813 &  $K=19$  & @ & 10 min  \\
6.29& 12.8 & 050904 &  $J=18$  & @ & 3 hr \\
5.6 & 12.6 & 060927 &  $I=16$  & @ & 2 min  \\
5.3 & 12.6 & 050814 &  $K=18$  & @ & 23 hr   \\
5.11& 12.5 & 060522 &  $R=21$  & @ & 1.5  hr   \\
\noalign{\smallskip}\hline\noalign{\smallskip}
\end{tabular} \label{relout:ta1}
\end{table}
\vskip -.425 truein
\end{center}

\subsection{SGRBs}

At \emph{Swift}'s launch, the greatest mystery of GRB
astronomy was the nature of short-duration,
hard-spectrum bursts.  Although more than 50 lGRBs
had afterglow detections, no afterglow had been found
for any sGRB.  Swift provided the first sGRB X-ray afterglow 
localization with GRB 050509B and  HETE-2 enabled the first 
optical afterglow detection with GRB 050709.  These two bursts, 
plus Swift observations of GRB 050824, led to a breakthrough i
n our understanding
\citep{2005Natur.437..851G,2006ApJ...638..354B,2005Natur.437..845F,2005Natur.437..855V,
2005Natur.437..859H,2005Natur.438..994B,2005Natur.438..988B}
of sGRBs. BAT has now detected 60 sGRBs,
most of which have XRT detections, and about one
third of which have host identifications or redshifts
(an additional two have been detected by \emph{HETE-2}, one by \emph{INTEGRAL},
and two by \emph{Fermi}/LAT).
We now have $\sim$50 sGRB localizations.

In stark contrast to long bursts, the evidence to date
on short bursts is that they can originate from regions
with low star formation rate. GRB 050509B and 050724 were
from elliptical galaxies with low current star formation
rates while GRB 050709 was from a region of a star forming
galaxy with no nebulosity or evidence of recent star
formation activity in that location.
Recent {\it HST} observations of locations
of sGRBs in their hosts reveal that sGRBs
trace the light distribution  of their hosts
while lGRBs are concentrated
in the brightest regions \citep{2010ApJ...708....9F}.
SGRBs are also  different from lGRBs in
that accompanying supernovae are not detected
for nearby events \citep{2006ApJ...638..354B,2005Natur.437..845F,
2005Natur.437..859H}. Taken together, these
results support the interpretation that short bursts are
associated with an old stellar population, and may arise
from mergers of compact binaries [i.e., double neutron star
or neutron star - black hole (NS-BH) binaries].

\subsection{GeV Emission}

{\it Fermi} was launched into low-Earth orbit in June 2008 and has
two primary high-energy detectors: the Large Area Telescope (LAT)
which operates between 20 MeV and $\ga300$ GeV, and the Gamma-Ray
Burst Monitor (GBM) which operates between  8 keV and 40 MeV. So far
the LAT has detected 24 GRBs; two were sGRBs, and nine showed
extended emission. The emission from GRB 090902B included a 34 GeV
photon. One of the most luminous to date has been GRB 080916C
\citep{2009Sci...323.1688A} at a redshift of 4.35. It had extended
emission (18 min) and exhibited a lag in LAT  energies with respect
to GBM. GRB 090510 is unique in being the only short burst with a
known redshift (0.903) showing  GeV emission. The lack of detectable
time delay between specific peaks in the light curves of  GRB 090510
at different energies leads to strong constraints on Lorentz
invariance \citep{2009Natur.462..331A}. Recent theoretical work
\citep{2010MNRAS.409..226K} on the {\it Fermi}/LAT detected GRBs
suggests that these may represent unusually powerful explosions with
Lorentz factors $\ga$10$^3$ in which the entire progenitor is
obliterated.  The simplest model, namely an external shock with
synchrotron emission, can be used to take the early values (at
$\sim$10$-10^2$ s) of the observed  high-energy emission and
successfully predict the much later values of the optical and X-ray
afterglow  (at $\sim$10$^5-10^6$ s).

\subsection{GRB Summary}

Recent progress in GRB research has been strongly motivated by
observational discoveries. To date, {\it Swift} has detected about
600 GRBs, {\it Fermi}/LAT 24. High redshift GRBs are illuminating
the properties of the high$-z$ universe and probing into the era of
re-ionization. {\it Swift} finds sGRBs in different environments
than lGRBs; also sGRBs are not accompanied by supernovae. The
accumulating evidence provides support for the NS-NS merger model.
Many GRBs have delayed onset of GeV emission, and more have extended
high energy emission. Interesting constraints on the Lorentz factors
associated with outflow, and Lorentz invariance violation, also come
from the synergism between GeV and lower energy observations. It is
not currently know for certain whether GRB jets are made  primarily
of baryons or Poynting flux, but momentum for the latter idea
appears to be gaining strength.

 \subsection{Jet Launching Processes}

There are currently two primary lines of thought regarding the
creation and propagation of jets in GRBs.

The baryonic jet model, whose roots can be traced back to the
elegant analytical solutions of a relativistic blast wave by
\citet{1976PhFl...19.1130B}, posits that a jet containing about a
Jupiter's mass worth of gas, $\sim0.001\msun$, is somehow launched
near the BH created by the collapsar with a Lorentz factor
$\Gamma\simeq 10-20$ \citep{2003ApJ...586..356Z}. The jet propagates
through the dense stellar envelope of the progenitor star where it
is focussed and compressionally heated. After breaking free of the
stellar surface, the thermal energy of the compressed jet is
translated into bulk kinetic motion, further accelerating the jet to
a Lorentz factor of several  hundred. Subsequent deceleration by the
circumstellar medium of the jet, which is then idealized as being
the fragment of a relativistic shell so that the  Blandford \& McKee
formalism can be brought to bear \citep{1998ApJ...497L..17S}, can
then be used to infer jet beaming angles from putative achromatic
``breaks'' in the GRB decay light curves
\citep{2001ApJ...562L..55F}.

The Poynting flux jet model, which has been gaining momentum in
recent years, traces its roots back to \citet{1977MNRAS.179..433B}
who considered the electromagnetic (EM) extraction of energy from
within the ergosphere of a Kerr BH. In the last few years workers
have developed sophisticated numerical codes  that calculate the 3D
evolution of gas and EM fields from the inner edge of accretion
disks onto spinning BHs, taking into account both  general
relativity and magnetohydrodynamics
\citep{2005ApJ...630L...5M,2007MNRAS.375..513M,2007MNRAS.375..531M}.
The accretion disk inner edge provides a natural collimating
surface. These workers find a baryonic zone-of-exclusion within the
jet, which effectively suppresses any baryonic component. For
numerical stability the numerical codes need to have some mass
within each  grid point, therefore a small trace amount of matter is
constantly added within grid points where the EM fields try to
exclude it. These are called ``floor'' models. Previous studies that
attempted to place constraints on the jet Lorentz factors, which
were based on the baryonic jet assumption, have been called into
question. The ramifications on the jet dynamics of having a
predominantly Poynting flux jet have not been developed yet in any
detail.

\begin{figure}[ht!]
\begin{center}
\includegraphics[height=100truemm]{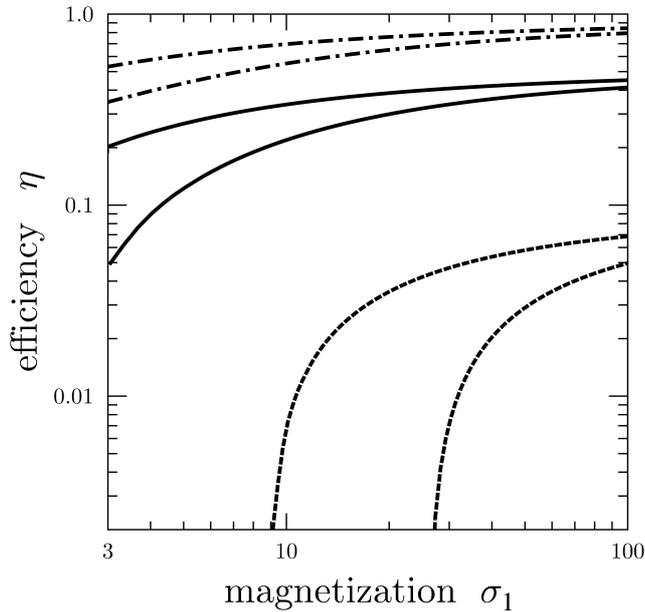}
\vskip -0.05 truein
\end{center}
 \caption{Simulated energy dissipation efficiency $\eta$ defined by Eq.\ref{eta} as
 a function of the  shell magnetization
parameter $\sigma$ for the inelastic collision of  two initially
cold magnetized shells of equal masses, but of different Lorentz
factors ($\Gamma_1 = 500$ and $\Gamma_2 = 1,000$). The key parameter
is the final magnetization of the hot merged shells $\sigma_{\rm f}$
was chosen to be 0.1 (dot-dashed curves), 1.0 (solid lines) and 10.0
(dashed lines). The two curves of the same line style in the Figure
differ by the assumed initial magnetization parameters. The top
curve for each line style corresponds to the case of equal initial
magnetization $\sigma_{\rm 1} = \sigma_{\rm 2}$, while the lower
curves of each type correspond to the fixed $\sigma_{\rm 2} = 0.1$.
The adiabatic index  $\gamma = 4/3$  was fixed for the matter in the
hot merged shell.}\label{eta1}
\end{figure}

\begin{figure}[ht!]
\begin{center}
\includegraphics[height=100truemm]{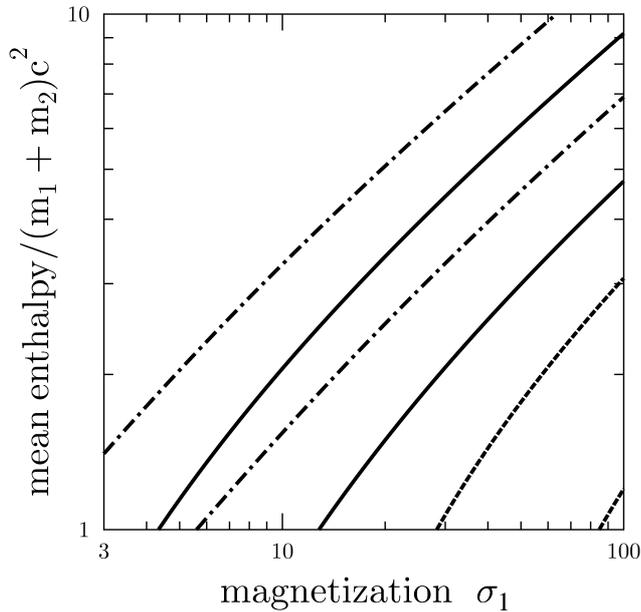}
\vskip -0.05 truein
\end{center}
 \caption{Simulated dimensionless mean enthalpy that is characterizing the mean Lorentz factor
 of the randomized particles
 in the rest frame of a hot merged shell. The curves are simulated
 for the same parameter sets as it is indicated in Figure
 \ref{eta1}.} \label{mean_gamma}
\end{figure}

\begin{figure}[ht!]
\begin{center}
\includegraphics[height=100truemm]{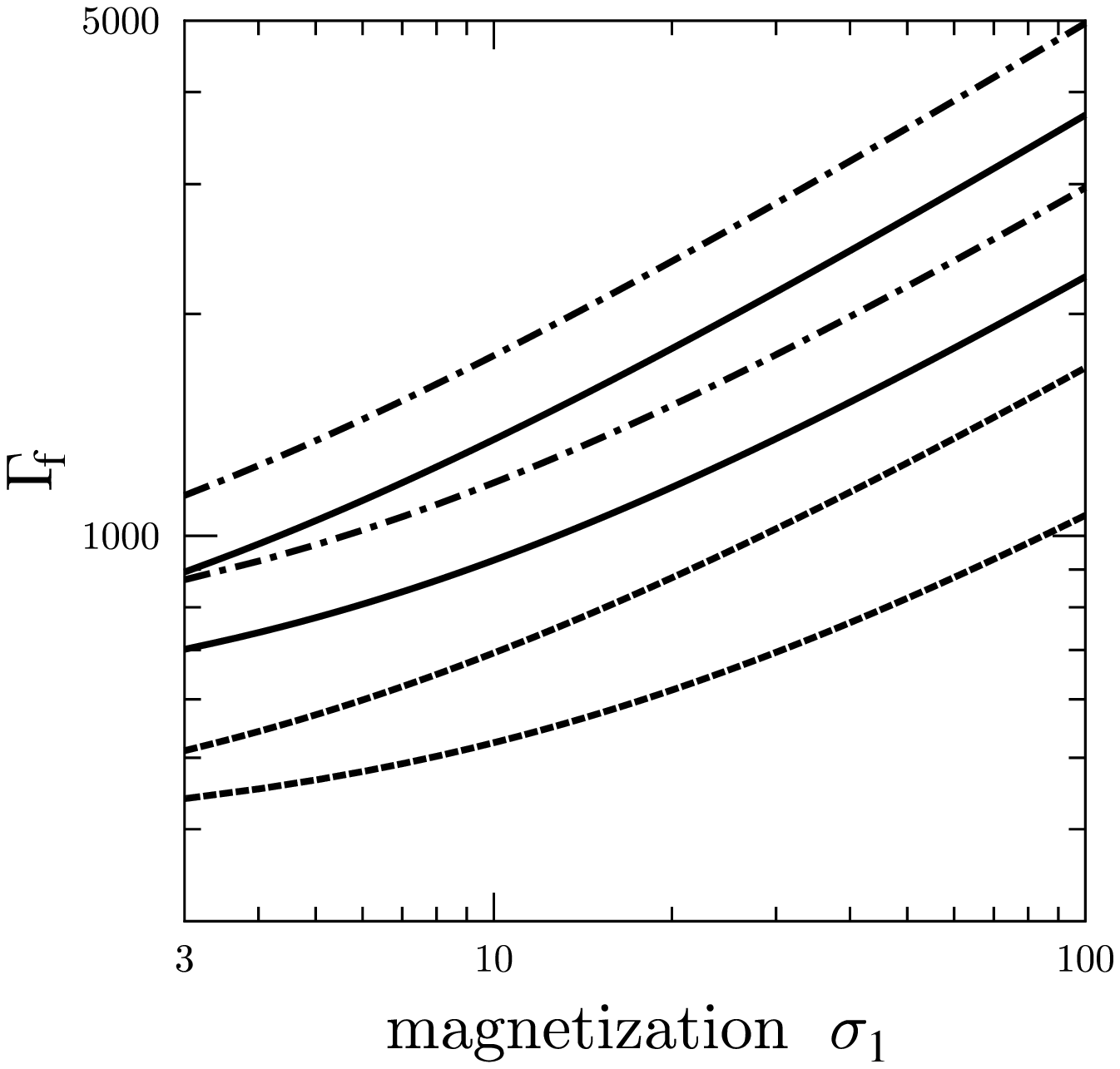}
\vskip -0.05 truein
\end{center}
 \caption{Simulated Lorentz factor $\Gamma_{\rm f}$ of a hot merged shell in the observer frame. The curves are simulated
 for the same parameter sets as it is indicated in Figure
 \ref{eta1}.} \label{gammaf}
\end{figure}

\section{Nonthermal Particle Acceleration in GRBs: challenges and perspectives}

Physical mechanisms of prompt emission in GRBs are still to be
established. There are fundamental questions of the powerful
relativistic outflow composition and matter vs. magnetic field
dominance to be addressed with both observations and advanced
models. The relativistic outflows may be different for two types of
GRB progenitors under consideration. Long and soft GRBs are most
likely connected to supernovae, while the short and hard GRBs are
possibly related to compact binary mergers.

Magnetized outflows from GRB engines of different nature have been
studied for some decades \citep[see,
e.g.][]{usov92,thompson94,mr97,lb04,zy11}. The outflows converting
the rotation power of a compact collapsar into a broad band of
radiation are likely models of GRBs and AGNs. The principal question
here is our understanding of the microscopic mechanisms of the
conversion of magnetic field energy into non-thermal particles and
the observed emission -- with or without shock formation.
Anisotropic striped wind with alternating magnetic polarity is
considered a favorable configuration to convert the magnetic energy
into the observed radiation of pulsar wind nebulae
\citep[e.g.][]{klp09,2011ApJ...726...75S}. An alternating magnetic
field configuration may occur in relativistic jets of GRBs. Magnetic
field reconnection demonstrated to produce electric fields that
accelerate particles in the Earth magnetosphere, solar flares, and
some laboratory plasma configurations \citep[see e.g.][]{yk10} was
also proposed as a plausible GRB model by
\citep[e.g.][]{sdd01,lb04,gs05,zy11,mu12}.

The most elaborated model of the origin of the GRB prompt emission
is, by now, the relativistic dissipative fireball model. There are a
number of alternative particle acceleration and radiation processes
within the relativistic dissipative fireball paradigm \citep[see
e.g.][]{rm94,piran04,mesz06,ma12}. Synchrotron and inverse Compton
radiation in the optically thin regions of the relativistic fireball
can be associated with the non-thermal electron/positron accelerated
either in the internal dissipation processes due to shocks or to the
flow magnetic field reconnections. Another potential component is
the photospheric emission (thermal or non-thermal) that is rather
rarely identified in the observed GRB emission. The prompt emission
light curves and spectra are generally in agreement with the
internal shock models. However, some potential problems of the
scenario are its efficiency and the lack of a bright photospheric
component observed in a few GRBs. We shall discuss now some general
features of the internal dissipation models with emphasis on
particle acceleration processes.

\subsection{Energetic Efficiency of Internal Dissipation Models: Shell Collisions in Jets}

An important issue of the internal dissipation scenario where the
energy carried out by multiple colliding shells of different
magnetization is the efficiency of the conversion of the outflow
power into the observed radiation. The variability of the central
engine (of a timescale $t_{\rm var}$) driving the relativistic
outflows of the mean Lorentz factor $\Gamma$, either matter or
Poynting dominated, can be modelled as a collision of
energy-containing shells. The model is considered to explain the main
features of the GRB prompt emission \citep[see
e.g.][]{piran04,mesz06}, as a vital alternative to the photospheric
models of GRBs. The dissipation region typically exists at the radii
about $r_{\rm diss} \sim ct_{\rm var}\Gamma^2$. In the case of
matter dominated jets (of low magnetization) the inner dissipation
occurs in the internal shocks while in the electromagnetically
(Poynting flux) dominated jets the magnetic field reconnection
effects are most likely crucial though shocks may also occur. The
microphysics of the dissipation in relativistic shocks as well as
modeling of the magnetic field reconnections are under intense
studies \citep[][]{yk10,bt11,2011ApJ...726...75S,mu12}. Realistic
models of a jet that would simulate the global RMHD dynamics and
simultaneously resolve the dissipative microphysical plasma
processes at much smaller spatial scales are not feasible at the
moment. However, simple multiple shell models of the internal
dissipation that just parameterize the magnetic field reconnection
effects are still rather useful and it is instructive to discuss
some of these models. A similar approach can be applied to other
relativistic outflows, like those of the AGN jets (see for a
discussion \S\ref{agn}) and of the pulsar wind nebulae.

To illustrate the effect of the outflow magnetization $\sigma =
B^2/4\pi \Gamma \rho c^2$ on the energy conversion into the observed
radiation it is instructive to use a simple two shell model
\citep[see e.g.][]{psm99,kumar99,dm98,zy11}. The shell collision may
result in dissipation of the magnetic energy due to reconnection and
turbulence cascade. Consider two shells of masses, Lorentz factors
and magnetization parameters [$((m_1, \Gamma_1,\sigma_{1})$ and
$(m_2, \Gamma_2), \sigma_{2})$], respectively, colliding
inelastically with the formation of a merged shell of $(m_{\rm f},
\Gamma_{\rm f}, \sigma_{\rm f})$. The internal energy $\delta'$
released in the rest frame of the merged shell (of the Lorentz
factor $\Gamma_{\rm f}$) is assumed to be in the form of either
thermal or non-thermal accelerated particles and radiation with
nearly isotropic distribution in the rest frame. Then, in the
observer frame the released energy is $\Delta E =\Gamma_{\rm f}\,
\delta'$. This leads to conversion of some amount of magnetic energy
into internal energy of the fluid, and then to radiation. Let us
envisage a picture where the two shells merge with a lower
magnetization parameter $\sigma_{\rm f}$ by the end of such an
inelastic collision. Energy conservation and momentum conservation
can be presented as
\begin{equation}
\Gamma_1\,\Psi_{1}\,m_1+ \Gamma_2\,\Psi_{2}\,m_2 = \Gamma_{\rm
f}\,\Psi_{\rm
f}\,(m_1+m_2+\delta')-\frac{\gamma-1}{\gamma\Gamma_{\rm
f}}\delta',~~~~\Psi_{\rm i}(\Gamma_{\rm i}) = 1 + \frac{2\Gamma_{\rm
i}^2-1}{2\Gamma_{\rm i}^2}\sigma_{\rm i}
\end{equation}
and
\begin{equation}
\Gamma_1 \beta_1 m_1(1+\sigma_{1}) + \Gamma_2 \beta_2
m_2(1+\sigma_{2}) =\Gamma_{\rm f}\, \beta_{\rm f}\,
(m_1+m_2+\delta')\, (1+\sigma_{\rm f}).
\end{equation}
Since the merged shell can not be considered as a cold one anymore
\begin{equation}
\delta'=\frac{(P'+\rho'_{\rm f})~V'}{c^{2}},
\end{equation}
where $\rho'_{\rm f}$ is the proper kinetic energy density (i.e. the
internal energy density with the rest energy density subtracted),
$V'$ is the shell volume, and $P'$ is the pressure in the rest frame
of the merged shell. Note that to calculate the energy released in
the rest frame of the merged shell $\delta'$ one should keep all of
the terms of the order of $\Gamma_{\rm i}^{-2}$. The expression for
$\Psi_{\rm i}(\Gamma_{\rm i})$ is exact for the case of the
transverse magnetic field in the rest frame of a shell, and it
accounts for the energy of the induced electric fields in the
observer frame.

In general, there are a few distinct components that contribute to
the pressure and the proper energy density (thermal and nonthermal
baryons, leptons, and photons). If for simplicity we describe these
as a single fluid with adiabatic index $\gamma$, then
\begin{equation}\label{eq_state}
 P'+\rho'_{\rm f}=\frac{\gamma P'}{\gamma-1},
\end{equation}
and one can resolve the energy-momentum conservation equations,
assuming the simple equation of state Eq.\ref{eq_state} to calculate
the Lorentz factor of the merged shell $\Gamma_{\rm f}$ and the
energy dissipation efficiency $\eta$ of the inelastic collision of
the two cold magnetized shells
 \begin{eqnarray}
 \eta & = & \frac{\Gamma_{\rm f} \delta'}{\Gamma_1\, \Psi_1 m_1+ \Gamma_2\,\Psi_2
 m_2}.
\label{eta}
 \end{eqnarray}

In Figure~\ref{eta1} we show the simulated energy dissipation
efficiency $\eta$ as a function of the shell magnetization parameter
$\sigma$ for the inelastic collision of two initially cold
magnetized shells of equal masses, but of different Lorentz factors.
We consider the GRB jet as a generic case and therefore choose
$\Gamma_1 = 500$ and $\Gamma_2 = 1,000$. The energy dissipation
efficiency $\eta$, the mean enthalpy of the hot matter in the merged
shell, that characterizes the mean Lorentz factor of the randomized
particles in the rest frame of the hot merged shell (shown in Figure
\ref{mean_gamma}), and the Lorentz factor $\Gamma_{\rm f}$ of the
hot merged shell in the observer frame, shown in Figure
\ref{gammaf}, are derived from the conservation laws and the
equation of state. The adiabatic index is fixed to $\gamma = 4/3$,
though in more accurate numerical simulations it depends on the the
the Lorentz factor $\Gamma_{\rm f}$. The key parameter here is the
final magnetization of the hot merged shell $\sigma_{\rm f}$ that is
determined by the currently poorly known rate of magnetic field
reconnection in the merging shells. The case of fast field
reconnection (and, therefore, efficient magnetic field dissipation)
is illustrated by the dot-dashed curves in Figure~\ref{eta1} where
$\sigma_{\rm f}$ = 0.1. The cases of lower magnetic field
dissipation are illustrated by the final magnetization parameters
$\sigma_{\rm f}$ = 1.0 (solid lines) and $\sigma_{\rm f}$ = 10.0
(dashed lines).

The dependence of the merged shell parameters on the initial
magnetization of the colliding shells is presented by two curves of
the same style that differ by the initial magnetization parameters.
The top curve for each type of line corresponds to colliding shells
of equal initial magnetization $\sigma_{\rm 1} = \sigma_{\rm 2}$.
The lower curves of each type correspond to the the case when the
fast shell of the Lorentz factor $\Gamma_2 = 1,000$ has low initial
magnetization $\sigma_{\rm 2}$ = 0.1.

It is clearly seen in Figure~\ref{eta1} that the dissipation
efficiency is higher in the case of the initially highly magnetized
shells with fast magnetic field reconnection resulted in low
$\sigma_{\rm f}$ of the merged shell (dot-dashed curves). Again, the
microscopical model of the reconnection rate in the complex flow is
still to be done to estimate the crucial value of the final
magnetization parameter $\sigma_{\rm f}$ of the merged shell
\citep[see e.g.][]{yk10}. To investigate the problem of shock
formation in the internal dissipation scenario, a microscopic
modeling of the collisionless shock formation and its structure in a
highly magnetized relativistic outflow is needed, and that is a
truly challenging task \citep[see e.g.][]{bt11,2011ApJ...726...75S}.
The problem of Fermi acceleration in transverse relativistic shocks
of different magnetization that are important to describe the
external shocks in the jets of GRBs and AGNs will be addressed in \S
\ref{rs}, here we concentrate on the internal dissipation models.

As it is seen in Figure~\ref{gammaf}, the high Lorentz factors
$\Gamma_{\rm f}> 1,000$ of a hot merged shell in the observer frame
can be achieved even for the incomplete magnetic field dissipation
in the merged shell (solid and dashed curves) and, therefore, it can
further catch up other slowly moving shells and merge with them
providing a chain dissipation process (Bykov \& Osipov 2012, in
press).

The mean Lorentz factors in Figure~\ref{mean_gamma} derived from the
conservation laws do not preclude a presence of a non-thermal (e.g.,
a piece-wise power-law) particle distribution, where some minor
fraction of particles can reach energies that are by some orders of
magnitude larger then the derived "thermodynamic" mean Lorentz
factor. Now we turn to discuss in brief possible particle
acceleration processes.

\subsection{Particle Acceleration in the Internal Dissipation
Models}\label{nonFP}

Relativistic turbulence produced by the internal dissipation in
shocks and magnetic reconnection in relativistic jets result in
acceleration processes occuring on both gyro time scale and on
longer comoving hydrodynamical time scales of the order of $l/c$.
The electric fields induced by turbulent motions of plasmas carrying
magnetic fields on different scales lead to statistical energy gains
of the superthermal charged particles and their wide-band radiation
\citep[see][]{bm96,Mizunoea10,Nishikawaea10,2011ApJ...726...75S,zy11,
2012MNRAS.421L..67B,muraseea12}.

For nonrelativistic MHD turbulence the particle energy gain over a
turbulent correlation length (or correlation time) is small, because
the induced electric field is smaller then the entrained magnetic
field. However, the distinctive feature of statistical acceleration
in the {\it relativistic} MHD turbulence and shocks on larger scales
expected in the flow-colliding regions, is the possibility of a
substantial particle energy gain over one correlation scale, because
the induced electric fields are no longer small. In this case the
standard Fokker-Planck approach cannot be used. Instead, it is
possible to calculate the energy spectra of nonthermal particles
within a special integro-differential equation which is a
generalization of the Fokker-Planck approach
\citep[see][]{bt93,bm96}.

Charged  particles interact with a wide spectrum of RMHD fields and
an internal shock ensemble produced by the colliding shells. In the
comoving frame, it is assumed that the fluctuations on all scales up
to $\sim \Delta$ (including the internal shock ensemble) are nearly
isotropic (in the latter case, it is enough if they are
forward-backward symmetric). The small mean free path $\lambda$ of
the superthermal particles leads to their isotropy in the frame of
the local bulk velocity fluctuations. The assumed statistical
isotropy of the bulk velocity fluctuations in the comoving frame of
the wind results then in a nearly isotropical particle distribution,
after averaging over the ensemble of strong fluctuations on scales
$\sim l$.

To calculate the spectrum of nonthermal leptons accelerated by an
ensemble of internal shocks and large-scale plasma motions in the
flow-colliding region (averaged over the statistical ensemble of large-scale field
fluctuations) we use a kinetic equation for the nearly-isotropic
distribution function $N=\gamma^2 F$, which takes into account the
non-Fokker-Planck behavior of the system \citep[see][]{bt93,bm96},
\begin{eqnarray}\label{kineq}
\frac {\partial F({\rm r},\xi,t)}{\partial t} = &  Q_i(\xi) +
  \int_{-\infty}^{\infty}
{\rm d} \xi_1 \; D_1 (\xi ~-~\xi_1 )\;\Delta F({\rm r},\xi_1,t) \nonumber\\
 & + \left(\frac {\partial^2 }{\partial \xi^2} +
    3\frac {\partial }{\partial \xi} \right)\int_{-\infty}^{\infty}
   {\rm d}\xi_1\;D_2(\xi~-~\xi_1)\:F({\rm r},\xi_1,t)
\end{eqnarray}Here $\xi =\ln(\gamma/\gamma_i)$, $\gamma_i$ is the Lorentz factor
of the injected particles, $Q_i(\xi) \propto \zeta c n l^2$ is the
rate of nonthermal particle injection, $n$ is the lepton number
density in the local flow comoving frame. The kernels of the
integral equation Eq.(\ref{kineq}) determining the spatial and
momentum diffusion are expressed through correlation functions which
describe the statistical properties of the large scale MHD
turbulence and the shock ensemble. Following the renormalization
approach, the Fourier transforms of the kernels $D_1^F(s)$ and
$D_2^F(s)$ are solutions of a transcendental algebraic system of
equations of the form $D_{1,2}^F = \Phi_{1,2} (D_1^F, D_2^F,s)$.
Here $s$ is a variable which is the Fourier conjugate of $\xi$.
Equation (\ref{kineq}) and the renormalization equations are valid
only for particles with sufficiently small mean free paths
$\lambda(\gamma) \ll \Delta$.

It is important to note that the solution of equation (\ref{kineq})
has a universal behavior, only weakly dependent on the complicated
details of the turbulent system. The stationary solution to
Eq.(\ref{kineq}) with a monoenergetic injection rate $Q_i$ has an
asymptotical behavior of a power-law form, $N \propto Q_i
\gamma^{-{\rm a}}$, where ${\rm a} = - 0.5 + [2.25 + \theta
D_1(0)D_2^{-1}(0)]^{0.5}$, and thus one may take $\theta \sim
(l/\Delta)^2$. For conditions typical of a developed RMHD
turbulence, the ratio of the rate of the scatterings to the
acceleration rate is $D_1(0)D_2^{-1}(0) < 1$, and for $\theta <1 $
one obtains ${\rm a} \sim 1$. This hard $\gamma^{-1}$ spectral
behavior arises because the acceleration time $\tau_a\sim l/c \sim
\alpha\Delta/c$ is much shorter than the escape time at the relevant
energies, $\tau_{esc}\sim\Delta^2/\kappa\sim\Delta^2/(l c) \sim
\Delta/(\alpha c)$. The power needed to produce such a spectrum of
nonthermal particles increases $\propto \gamma_{max}$, so it is
important to understand its temporal evolution.

In the test particle limit, where the backreaction of the
accelerated leptons onto the energy-containing bulk motions is
small, we have $N(\gamma,t) \propto \zeta n \gamma^{-1}$ for $\gamma
\leq \gamma_{\star}(t)$, where $ \gamma_{\star}(t) = \gamma_{i}
\exp(t/\tau^h_a)$ and
\begin{equation}
\tau^h_a \propto l/c \sim \alpha (\Delta /c)~,
\end{equation}
is the typical hydrodynamical acceleration timescale \citep[see,
e.g.][]{bt93}, with $\gamma_{i}\sim$ few, $\alpha < 1$, and the
comoving width of the region energized by shocks equal to $\Delta$.
From the energy balance equation, when the value $\gamma_{\star}(t)
\sim \gamma_i~ m_p/m_e~\epsilon~\zeta^{-1}$ is reached, the growth
must saturate, and the resulting spectrum consists of two branches.
One of these is the hard spectrum $N(\gamma)  \sim \zeta n
\gamma^{-1}$, for $\gamma \leq \gamma_{\star}$, where
\begin{equation}
\gamma_{\star} \sim \gamma_i~m_p/m_e~ \epsilon~\zeta^{-1}~.
\end{equation}
For the typical scales of the considered problem $\gamma_i \sim 1$
and $\zeta\sim 10^{-3}$, so $\gamma_\star\sim 10^{5}$ (but it could
be even larger since $\epsilon \sim 1$ for large scale plasma
motions). In the regime of very efficient particle acceleration
where the backreaction of the accelerated leptons on the
energy-containing bulk motions is substantial, one should use a
non-linear approach. In that case the particle distribution
experience fast temporal evolution \citep[see e.g.][]{bykov01,fm10}.

The multiwavelength observations discussed above can be used to
constrain the characteristics of relativistic turbulence and
particle spectra evolution. Measurements of polarization of prompt
gamma-ray emission of GRBs would be very informative. Polarization
measurements provide an additional deep insight into the nature of
acceleration/radiation processes. By now only a few instruments are
available for this kind of measurement at gamma-ray energies.
Gamma-ray bursts (GRBs) are very promising candidates for
polarimetry due to their large flux over limited time intervals,
maximizing the available signal-to-noise ratio. To date, a few
polarization measurements have been reported, all claiming a high
degree of polarization in the prompt emission of GRBs, but with
rather low statistical evidence yet. \citet[][]{goetzea09} used the
IBIS telescope aboard the INTEGRAL space laboratory to measure the
polarization of the prompt gamma-ray emission of the long and bright
GRB 041219A in the 200--800 keV energy band. They found a
variable degree of polarization ranging from less than 4\%
over the first peak to 43\%-25\% for the whole second peak. Time-resolved
analysis of both peaks indicated a high degree of
polarization, and the small average polarization in the first peak
can be explained by the rapid variations observed in the
polarization angle and degree. The results by \citet[][]{goetzea09}
are consistent with different models for the prompt emission of GRBs
at these energies, but they favor synchrotron radiation from a
relativistic outflow with a magnetic field, which is coherent on an
angular size comparable with the angular size of the emitting
region. Recently, \citet[][]{Yonetokuea11} reported the polarization
measurement of the prompt gamma-ray emission of GRB 100826A with the
Gamma-Ray Burst Polarimeter (GAP) aboard the small solar-sail
demonstrator IKAROS. They detected the change of polarization angle
during the prompt emission, and the average polarization degree  of
27$\pm$11\%. Polarization measurements are a powerful tool to
constrain the GRB radiation mechanisms and the magnetic field
structure in the photon emitting regions.

\section{Active Galactic Nuclei}\label{agn}
\subsection{Studies of Particle Acceleration in AGN Jets}
The relativistic outflows of Active Galactic Nuclei (AGNs) are
well-studied particle accelerators. The AGN jet phenomenon spans many orders of magnitudes:
jets originate on sub-pc scales close to supermassive black holes with typical masses between
a few million and a few billion solar masses and can propagate over many hundred kpc
to feed giant hot-spot complexes and radio lobes.
One of the most remarkable properties of jets is that they dissipate little energy
while transporting vast amounts of energy and momentum over large distances.
However, jets are not dissipation-free neither one sub-pc scales nor on kpc-scales:
they do accelerate electrons and maybe also protons to high energies.
The high-energy particles interact with magnetic fields, photon fields,
and/or ambient matter and make AGNs some of the brightest extragalactic sources of
continuum emission across the electromagnetic spectrum.
There are many reasons to study particle acceleration in AGNs: we would like to
explain the observed electromagnetic radiation in order to constrain the composition
and structure of jets. The composition and structure constrain the processes of
AGN accretion and jet formation, acceleration and collimation.
The same studies can enhance our understanding of AGN feedback, i.\ e.\ how AGNs
interact with their hosts and decisively impact their evolution.
Studies of AGNs also allow us to study the particle acceleration
mechanisms. A good (“microscopic”) understanding of the dissipation
processes in jets is instrumental for addressing the larger
questions mentioned before.

In the following we discuss two particle acceleration sites:
the “blazar zone” less than a pc away from the supermassive black hole,
and jet particle acceleration by kpc-scale jets.
The processes at work in these two well-separated regions have recently
received a lot of attention – mainly because of spectacular jet
images from the VLBA, HST, and Chandra, and spectacular multiwavelength
observations with the RXTE, Suzaku, and Swift, X-ray, and Fermi, H.E.S.S.,
MAGIC, and VERITAS $\gamma$-ray telescopes.
\subsection{Studies of Particle Acceleration Processes in the Blazar Zone}
The cores of the ``blazar class'' of AGNs are bright sources of spatially
unresolved continuum emission. In the case of BL Lac objects, emission and
absorption lines are absent or weak. In the case of their more powerful siblings,
Flat Spectrum Radio Quasars (FSRQs), the spectra exhibit emission and
absorption lines, which afford additional diagnostics concerning the
mass of the central black hole, the accretion power, and the redshift of the source.
Blazars are sources with jets closely aligned with the line of sight.
The relativistic propagation of the jet plasma and the close alignment
of the jet with the line of sight lead to relativistic boosting of the emission.
Blazar Spectral Energy Distributions (SEDs) show evidence for two broad peaks –
presumably coming from synchrotron and inverse Compton emission from the same
electron population. Observations of the BL Lac Mrk 501 revealed emission up
to 16 TeV \citep{Ahar:99}, giving direct proof that AGN jets accelerate
particles to $\gg$ 1 TeV energies. AGNs may accelerate particles to much
higher energies, maybe even to ultra-high energies \citep[e.g.][]{Kach:10},
see however \cite{2009JCAP...11..009L}
for a detailed discussion of acceleration power in these sources.

Two scenarios are commonly invoked to explain the observed emission:
(i) the jet is initially Poynting flux dominated and accelerates particles
in magnetic reconnection events \citep[e.g.][and references therein]{Gian:10};
(ii) the jet is either particle energy dominated right from the start, or,
it is initially Poynting flux dominated and some unknown mechanism
converts the energy from Poynting flux into bulk motion energy,
and the particle dominated jets subsequently accelerate particles at shocks.

Blazar studies of particle acceleration benefit from the fact that the
broad-band SEDs oft he sources evolve on short time scales:
X-ray and gamma-ray flares with ~1 min durations have been reported.
It thus is possible to track the temporal evolution of the particle energy spectra.
Modeling of sequences of broad-band SEDs has shown that the jets are highly
relativistic with bulk Lorentz factors of $\sim$50 or even higher \citep[e.g.][]{Kraw:01}.
The simplest types of models – synchrotron self-Compton models - indicate that the
particle energy density dominates strongly over the magnetic field energy density
in the blazar zone \citep[e.g.][]{Kraw:02,Acci:11}.
These results clearly favor shock acceleration over magnetic reconnection.
The data and theoretical studies (particle-in-cell (PIC) simulations and instability analyses)
favor mildly relativistic shocks internal to the jets over highly relativistic external shocks:
the former have faster downstream plasmas commensurate with the high inferred bulk Lorentz
factors of the emitting plasma, and are less susceptible to the strong suppression of the
formation of upstream scattering centers by even a weak plasma magnetization
than their highly relativistic counterparts \citep[e.g.][]{Lemo:10,Siro:09}.

X-ray and very high-energy $\gamma$-ray observatories have recently succeeded to
sample the temporal evolution of the X-ray and γ-ray energy spectra with delicate accuracy.
Although observers organized a large number of observation campaigns with the objective
to find the flux vs.\ spectral index evolution patterns predicted by simple
acceleration theory \citep[e.g.][]{Kirk:98}, the observations revealed
rather erratic evolutions \citep[e.g.][and references therein]{Taka:00,Gars:10}.
Unfortunately, observations have not yet succeeded to determine unambiguously
where exactly the blazar emission originates. For the radio galaxy M87 –
possibly a misaligned blazar  – the observation of temporally coincident radio,
X-ray, and $\gamma$-ray flare indicates that the emission comes from $<$100
Schwarzschild radii of the supermassive black hole \citep{Acci:09}.

\subsection{Studies of Particle Acceleration Processes of kpc-scale Radio, Optical, and X-Ray Jets}
The VLBA, HST, and Chandra telescopes are delivering images of the kpc-jets of an ever increasing
number of radio galaxies. The images in the radio, optical, and X-ray bands can be used to infer
complementary information about the emitting particle populations.
Radio and optical polarimetry give additional clues about the orientation of the magnetic field
in the bright jet regions.

A recent somewhat surprising result was the detection of a large number of X-ray bright
kpc-jets with Chandra \citep[see the review by][and references therein]{Harr:06}.
In the case of powerful Fanaroff-Riley (FR) type II sources (like PKS 0637-752)
the combined radio, optical and X-ray energy spectra showed unambiguous evidence
for two distinct emission components. Presently two models are commonly invoked
to explain the “second” X-ray bright component: (i) inverse Compton scattering
of Cosmic Microwave Background photons \citep{Tave:00,Celo:01},
(ii) a second synchrotron component \citep{Harr:02}. The first model requires jet plasma
moving with large bulk Lorentz factors ($\Gamma\sim$10) at kpc-distances from the
central engine. The relativistic motion boosts the CMB photon energy density in
the reference frame of the emitting plasma by a factor of $\Gamma^2$, and the
mean photon energy by a factor of $\Gamma$. These two effects make it possible
to explain the observed X-ray emission with electrons with modest electron
Lorentz factors on the order of $\gamma =$~100.
The second model requires much higher Lorentz factors: assuming equipartition
magnetic fields, electrons with Lorentz factors $\gamma\sim 10^7$ are needed to
explain the X-ray emission. Both models have severe shortcomings. For example,
in the first model the long cooling times of the low-$\gamma$ electrons are at
odds with the well-defined knotty structure of some X-ray bright jets.
A weakness of the second model is that there is not yet a good explanation for
the existence of two distinct electron populations with very different
spectral properties. For lower-power FR-I-type sources like M 87, Cen A, or 3C 31,
the combined radio, optical and X-ray energy spectra are consistent with a
synchrotron-origin of the X-ray emission from a single population of electrons.
For an equipartition magnetic fields of $B\sim$100~$\mu$G, the X-ray emitting
electrons have Lorentz factors $\gamma\sim 10^7$, and radiative synchrotron
cooling times of a few years. The X-ray emitting electrons thus do not move
far from their acceleration sites before they loose their energy, and the X-ray
brightness profiles tracks the acceleration of the high-energy electrons.
The diffuse appearance of some jets implies quasi-continuous acceleration.

A few radio galaxy jets have been studied with the Hubble Space Telescope
giving not only high-resolution images of the optical brightness but also
of the optical polarization. The magnetic field probed by the optically emitting
electrons seems to be aligned parallel to the jet flow for the most part.
However, upstream of the brightness maxima, the field are perpendicular to
the jet flow. The radio and optical polarization behavior differs,
indicating that the emission at different wavelengths samples different
regions of the jet \citep{Perl:99}.

\section{Acceleration to ultra-high-energies}
\subsection{Some properties of UHECRs}

One of the 11 fundamental science questions
for the 21st century listed in the final report of the 2002 Decadal Review
\citep{turner} is the nature of cosmic rays. The detection of cosmic rays at
ultra-high energies (UHECR) dates back to the early Sixties, but only during the last
20 years detectors of sufficiently large size have become operational, that the
origin of UHECRs
can be addressed \citep[for a review see][]{2011ARA&A..49..119K}. UHECRs are indirectly
detected by observing the airshowers they trigger in the atmosphere. One can look for
fluorescence emission or other radiation produced high in the atmosphere, or, alternatively, one registers the passage of secondary particle in charged-particle
detectors on the ground. The AUGER observatory \citep{2004NIMPA.523...50A}
and the Telescope Array \citep{2008NuPhS.175..221K} combine both techniques.

The main observables used to infer the properties of UHECRs are the anisotropy, the composition, and the spectrum. The composition is difficult to determine, because the appearance of giant airshowers can only be modeled with particle-physics event generators that involve extrapolations of behaviour observed in accelerator experiments, for the
CoM energy of an arbitrary nucleus of 10~EeV energy with a nitrogen nucleus at rest is far higher than that achievable with even the largest man-made accelerator, the LHC at CERN. Considerable systematic uncertainty thus overshadows attempts to study the composition. It appears that between 1~PeV and 0.1~EeV we observe a trend from
a predominance of light particles to heavy nuclei. Around 1~EeV, the composition is
light again \citep{2010PhRvL.104p1101A,2010PhRvL.104i1101A}. Above 1~EeV, Auger
observes a transition to heavier particles that is not see with other
experiments at this time, possibly on account of statistics.

The anisotropy is low around 1~EeV, where upper limits near 1\% have been published for the sidereal dipole anisotropy \citep{2011APh....34..627P}. At higher energies above 57 EeV, for which little deflection would be expected, if the primary particles were protons, a correlation is observed between the arrival direction of particles and
certain types of nearby AGN which in the end are proxies of the matter distribution
within $\sim$75 Mpc from us \citep{2008APh....29..188P,2010APh....34..314T}.
The distance limitation is expected because nuclei at these energies undergo photodisintegration and photomeson
production that provide losses on corresponding time scales, leading to the so-called
GZK cut off.

Whereas cosmic rays approximately obey a power-law spectrum with index
$s\simeq 2.7$ ($dN/dE\propto E^{-s}$) below the so-called knee in the spectrum at 3 PeV, the spectrum of UHECRs is soft between a few PeV and 3~EeV with a power-law index
$s\simeq 3$ \citep{2009APh....31...86A}. At 3~EeV the spectrum hardens to $s\simeq 2.6$, a feature known as the ankle. Above 30~EeV one observes a flux suppression that has been identified with the GZK cut off \citep{2010PhLB..685..239A}
A recent proposal \citep{Aloisio11}
interprets those features as a proton cut off around
$10^{18}eV$ and another one around $3 \times 10^{19} eV$ associated
with iron nuclei\footnote{This scenario implicitly postulates a very large
proton-to-helium ratio in the source}.

\subsection{Implications of the maximum energy}\label{ssec:maxE}
Cosmic rays at energies below
1 PeV are almost certainly
galactic in origin, and those at energies above 10 EeV are most likely extragalactic,
but considerable uncertainty exists at intermediate energies.
It is unclear at what energy the
local cosmic rays turn from being predominantly galactic to being mostly extragalactic.
The relevance of this uncertainty for modeling the sources of cosmic rays is obvious: if the particles in the energy band above the knee at a few PeV, or above the iron knee at $\sim 10^{17}$~eV, are extragalactic, then considerable
finetuning is required in matching the galactic and extragalactic components, because
the spectrum softens at the knee. On the other hand, if cosmic rays
up to a few EeV, i.e. up to the ankle in the spectrum, are galactic,
then no such finetuning is required, but we need
to identify the sources of EeV-band cosmic rays with objects present in the Galaxy, e.g.
supernova remnants (SNR), pulsars, etc. This can be difficult, not because the source in
question would not accelerate particles to high energies, which in fact we observe
happening in SNR and pulsars, but because it is questionable that EeV energies can be reached.

In fact, for typical
interstellar magnetic field values, SNR shock fronts can hardly accelerate cosmic
rays to a PeV \citep{lc1,lc2}.
Particle confinement near the shock is supported by self-generated magnetic turbulence ahead of and behind the shock.
Various plasma instabilities driven by cosmic rays can
contribute to excite the turbulence to high levels, although which
dominates remains an active topic of research. In the case of
SNRs, that which has received most attention so far
is the so-called streaming instability seeded by the cosmic-ray
net current \citep[e.g.]{W74,S75,A83,lb00,bl01}, and more recently
its non-resonant counterpart \citep{Bell04,Bell05,Pel06}. In
contrast, relativistic shocks operating at the interface between
AGN/GRB flows and the surrounding medium reveal a short precursor,
which restricts the plasma instabilities to small scale modes
\citep{1999ApJ...526..697M,ps00,pls02,2006PPCF...48.1741R,
2009MNRAS.393..587P,Lemo:10,2011MNRAS.417.1148L},
as discussed in detail in the following.
Clearly, the amplitude of the turbulence sets up the pitch-angle scattering
frequency and thus the acceleration rate \citep{md01}. In addition, it also sets
the scale for the maximum energy, to which a remnant may accelerate particles. Although analytical and
numerical estimates suggest that cosmic rays can very efficiently drive
magnetic turbulence ahead of the shock \citep[e.g.][]{bykov05},
so the turbulent magnetic field may be much larger
than the homogeneous interstellar field \citep{lb00,bl01}, large
increases in the magnetic field strength do not necessarily translate into a significant increase in
the maximum particle energy \citep{veb06}.

For relativistic sources such as AGN or GRB, relativistic shock acceleration can be invoked, but even
there certain limitations arise
\citep[e.g.][]{Gal99,Ach01,2009MNRAS.393..587P,LP11b,ep11}.
To be shock accelerated, a particle that has crossed the shock toward
the upstream must be overtaken again by the shock. Assuming the shock
moves at Lorentz factor $\Gamma_S$, the particle must have been
deflected (by gyration or scattering) through an angle $\Delta
\theta\gtrsim 1/\beta_S\Gamma_S$ while residing upstream. This
deflection must be accomplished within a time $\Delta t$ at least of
order $R_S/\beta_S c$, where $R_S$ is the shock radius at which the
shock once again overtakes the particle.  The factor $1/\beta_S$ in
$\Delta\theta$ arises for subrelativistic shocks on account of the
small incremental energy gain per shock crossing. The particle must
cross the shock $\sim 1/\beta_S$ times to double its energy, with each
crossing requiring at least a significant fraction of a gyroperiod.

The fastest possible deflection is provided by undisturbed gyration in
magnetic field oriented perpendicular to the shock normal, for which
the angle between particle momentum and shock normal, $\theta$,
increases linearly with time, and the deflection rate $\Delta
\theta/\Delta t$ must then obey $\Delta \theta/\Delta t=\beta c/r_g
\gtrsim c/(R_S\,\Gamma_S)$. This inequality sets a maximum energy
$E_{\rm max}$ to which a particle can be accelerated, because the rate
of change of angle presumably decreases with particle energy.  As
$r_g= pc/ZeB= \beta E/ZeB$, this corresponds to a maximum energy, in
the limit of relativistic particles, of
\begin{equation}
E_{\rm max}= Ze\,B\,R_S\, \Gamma_S
\label{emax1}
\end{equation}
Such regular deflection only occurs in a magnetic field that is
coherent over scales larger than the path length of the particle. In
the short precursor of a relativistic shock, this restricts $B$ to
the background, undisturbed magnetic-field value, and it thus limits
the maximal energy to a rather small value for typical interstellar-medium
conditions. To be noted is that Equation~\ref{emax1} may provide a rather
academic limit, because in a relativistic shock a perpendicular magnetic field in
the downstream region renders acceleration very inefficient.
Also note that the magnetic-field strength, $B$, is supposed to be measured in the upstream frame of the shock, i.e. it is the ambient field in the source frame in,
e.g., an extrenal shock of a GRB.

Note that scatter-free gyration cannot in general confine a CR particle to a subrelativistic
blast wave in all three dimensions. The particle generally drifts off to the
side after gaining the potential difference $\beta_S\,E\,B\,R_S$ in energy. Some scattering is required which will reduce $E_{\rm max}$.
Also note that we have neglected both adiabatic losses and drift to the periphery of the
shock front and assumed that being overtaken by a spherical blast wave is sufficient for
further acceleration.

For relativistic shocks, escape through the lateral boundaries
does not provide a stringent constraint on the maximal acceleration
energy unless sideways expansion of the blast takes place: as viewed
in the shock front rest frame, the particle is confined if its
gyration radius $r_{g,0\vert\rm sh} < R_\perp$, with $R_\perp$ the
lateral extension of the shock front. Since $r_{g,0\vert\rm
sh}\simeq r_{g,0}/\Gamma_S^2$, with $r_g$ the upstream gyroradius
in the background field, confinement leads to $E_{\rm max}<
\Gamma_S^2R_\perp e B$, which is not as restrictive as the previous
expression if $R_\perp > R_S/\Gamma_S$. In that limit, sideways
expansion of the blast is negligible and the overall dynamics
resembles that of a spherical blast wave.

What of often invoked magnetic-field amplification by cosmic-ray induced instabilities? The growth of plasma instabilities in the precursors of shocks is 
inevitable, and therefore in a realistic situation we cannot expect undisturbed 
gyration in perpendicular magnetic-field. In fact, random scattering is required if the large-scale magnetic field is oriented parallel to the shock,
because otherwise particles could not return to the shock. Therefore, 
random scattering in small-scale fields will make acceleration at parallel shocks faster, and thereby increase the maximum energy, in particular at nonrelativistic
shocks. 

The scattering mean free path can be written as
$\lambda\sim c/D_{\theta\theta}$, where the angular diffusion
coefficient is given by $D_{\theta\theta}=\delta \theta ^2/\delta t
\sim \left(eB_{rms}/\beta \Gamma mc\right)^2 l/\beta c$ where $\delta t \sim
l/\beta c$ is the scattering coherence time, over which the particle
scatters by an angle $\delta \theta$, and $l$ is the coherence
length of the magnetic field \citep[e.g.][]{ep11,Plotnikov11}.
At sub-relativistic shock waves, one must now impose
$r_g^2/l\le\beta_S R_S$, with $r_g= pc/ZeB_{\rm rms}$. For
relativistic shocks, the condition for the particle to suffer a
rms deflection $1/\Gamma_S$ over a timescale $R_S/c$ reads
$r_g^2/l\le\Gamma_S^2 R_S$, so that these two equations can be
combined into
\begin{equation}
E_{\rm max}\le Ze\,B_{\rm rms}\,\left(\beta_S l R_S\right)^{\frac{1}{2}}\, \Gamma_S \label{emax2}
\end{equation}
which is less than the previous expression when $l\ll r_g$. {Such
  small-scale fields are expected in the precursor of relativistic
  shocks, which cannot exceed $r_{g,0}/\Gamma_s^3$, although there is
  then an ambiguity related to the reference frame of the small-scale
  magnetic inhomogeneities; for simplicity, we have assumed here that
  these magnetic inhomogeneities are at rest in the upstream plasma.}
Thus, provided some large-scale perpendicular magnetic field exists,
simply tangling the field on small scales, $l$, does not necessarily
raise $E_{\rm max}$. Magnetic-field amplification enhances the maximum
energy only if it increases $B_{rms}^2\,l$.  Note that the expression
$E_{\rm max}= Ze\,B\,R_S$, often taken from Figure 1 of
\citet{hillas}, is consistent with equations (\ref{emax1}) or
(\ref{emax2}) only if $\beta_s $ and $\Gamma_s$ are both of order
unity. We stress again that the limits described here may not be
reached at a real shock. Leakage from the precursor and the conditions
downstream must also be considered when evaluating the maximum energy
and the acceleration efficiency.

Equations ~\ref{emax1} and \ref{emax2} suggest that the sources of
UHECRs are likely systems involving relatistic shocks.

\subsection{Sources of UHECRs}\label{sec:UHECR}

Besides reaching the required particle energy, the sources of UHECR must also be powerful
enough to provide the source luminosity needed to sustain the local flux of UHECR.
Possible source candidates of UHECRs are active galactic nuclei (AGNs)
\citep{s1,1990PThPh..83.1071T,1993A&A...272..161R,2009PhRvD..80l3018P},
clusters of galaxies
\citep{s2},
Magnetars
\citep{s3,2009PhRvD..79j3001M},
and gamma-ray bursts (GRBs)
\citep{s4,1995ApJ...453..883V,2006ApJ...651L...5M}.
Depending on the model, these sources may also dominate the energy range around $10^{18}\ {\rm eV}$
\citep{2006PhRvD..74d3005B,2007PhRvD..76h3009W,s5}.

One can constrain the acceleration capabilities of various sources
through the magnetic luminosity of these
sources~\citep[e.g.]{Nea95,W05,2009JCAP...11..009L}{, as
follows}. {Let us assume} that acceleration takes place in
an outflow at radius $r$ moving with possibly relativistic velocity
$\beta$ (and Lorentz factor $\Gamma$) towards the observer, so as to
benefit from Lorentz boosting. { We assume that acceleration
proceeds with an acceleration timescale $t_{\rm acc}\equiv{\cal
A}r_g/c$ in the comoving frame}, with ${\cal A}>1$.{ Then
the maximal energy at acceleration is at least bounded by the
condition $t_{\rm acc}<r/(\Gamma\beta c)$, which means that the
acceleration timescale must be shorter than the comoving age of the
outflow. This limit can be rewritten in terms of the maximal energy
in the observer frame, $E_{\rm max}$ and in terms of the magnetic
luminosity of the source, $L_B\equiv r^2\Theta^2\Gamma^2\beta c
B^2/4$ as calculated in the source rest frame in terms of the jet
half opening angle $\Theta$ and comoving magnetic-field strength
$B$:}
\begin{equation}
E_{\rm max}\lesssim 10^{20}\,{\rm eV}\,{\cal A}^{-1}\Gamma^{-1}\Theta^{-1}\beta^{-3/2}\,Z\,L_{B,45}^{1/2}\label{eq:LB}
\end{equation}
with $L_{B,45}=L_B/10^{45}\,$erg/s. One can check that this bound
remains robust in the small $\Theta$ limit, {meaning
$\Theta\Gamma\rightarrow 0$ for which } side escape becomes
important, and in the small $\beta$ limit. This bound indicates that
rather extraordinary luminosities are required to accelerate particles
to ultra-high energies, under rather general conditions, although the
bound depends on the charge of the particle. For instance, if one
derives the magnetic luminosity of blazars through a leptonic
modelling of the spectral energy distributions, one concludes that
only the rare flat spectrum radio quasars with jet powers
$\gtrsim10^{44}\dots 10^{46}$~erg/s can accelerate protons to $\sim
10^{20}\,$eV, while other Bl Lac and TeV blazars (FR~I analogs) with
jet powers $\sim10^{40}\dots 10^{44}$~erg/s appear limited to $\sim
10^{18}-10^{19}\,$eV \citep{2009JCAP...11..009L}. From this point of
view, more compact sources such as GRBs and magnetars appear
favored. For instance, a GRB of apparent isotropic luminosity
$10^{52}$erg/s with $\Gamma\sim 100$ may produce particles with energy
as high as $Z \times 10^{21}$~eV for a magnetic conversion factor
$\xi_B=0.01$.  As discussed in Sec.\ref{sec:acc}, mildly or
sub-relativistic shocks in a relativistic flow are more efficient
accelerators of protons than ultra-relativistic shocks and are
excellent candidates for being sources of UHECRs, owing to the
magnetic-field amplification at shocks.

The paucity of FR~2 radio-galaxies in the GZK sphere (radius $\sim
100$Mpc) capable of accelerating protons to ultra-high energies might
be compensated by the acceleration of heavier nuclei in the less
powerful and more numerous FR~I radio-galaxies. In particular, Ptuskin
and collaborators have shown that if radio-galaxies inject of light to
heavy elements with a rigidity dependent maximal energy following
Eq.~(\ref{eq:LB}), $L_B$ being related to the radio luminosity,
accounting for the radio-luminosity function, one could explain rather
satisfactorily the observed spectrum \citep{Ptuskin11}. It is also
intriguing that the Pierre Auger Observatory reports an excess of
events in the direction of the nearby radio-galaxy Cen~A (although
this latter happens to lie in front of one of the largest
concentrations of matter in the GZK sphere, the Centaurus
supercluster). However, it would be very difficult to understand the
observed pattern of anisotropy if one assumes that the highest energy
particles are heavier than hydrogen in such scenarios
\citep{2009JCAP...11..009L}.

Besides the actual source physics, the
evolution of sources, the number of accelerators within a source
\citep{s6},
and the variation of source properties
\citep{s7,2008ApJ...684L..69B}
will also shape the local spectrum of UHECRs.

Many properties of UHECRs can be impacted by their propagation in intergalactic
space, such as their composition through photo-desintegration or their spectrum through
cascading via photo-meson and photo-pair production, but it is difficult to disentangle
the propagation effects from the results of physical processes operating inside
the sources of these particles.

Estimating
the source luminosity using observed quantities and the known population statistics of the sources in
question is subject to considerable uncertainties. As an example, \citet{egp10} have recently estimated the
local UHECR source luminosity, assuming all particles above the ankle at 4 EeV are extragalactic, and
compared that with the observed gamma-ray production rate of all GRB. In contrast to earlier studies
\citep{wax2004,ld07}, not the MeV-band gamma-ray fluence was used, which likely represents a thermal pool,
but the GeV-band emission observed with Fermi-LAT, which measures the non-thermal tail
of the energy distribution in the GRB primary charged particles, which, if hadronic, is the part that
could contribute to the UHECR flux. It turns out that the UHECR source luminosity is more than a hundred
times higher than the total GeV-band photon output, which places severe constraints on UHECR
models involving GRB. If one posits that the Galactic to extra-Galactic transition takes place at
$\sim 10^{19}\,$eV and the MeV gamma-ray fluence traces the
nonthermal particle population, the particle output of GRBs is 
more commensurate with their photon output
\citep{2010arXiv1010.5007W}.

For each source class, one can also estimate
a luminosity function, that is the differential source density needed to integrate
the contribution of the sources over cosmological redshift. While only nearby
($\lesssim 200$~Mpc) sources may actually contribute to locally observable
GZK-scale UHECRs,
the interaction products of the particles from all more distant sources will
feed a cascade of energy that is eventually observable as a component of the extragalactic
gamma-ray and neutrino background, which are two other cosmic messengers that are
complementary to the charged particles.
The former has been recently measured with
unprecedented sensitivity up to 100~GeV
\citep{s8}, and thus provides
invaluable constraints on, e.g., the so-called dip models, which assume that essentially all particle above about 1~EeV are protons. The redshift of the onset of photopair production with the CMB would then naturally lead to an ankle at the energy where it is indeed observed \citep{2006PhRvD..74d3005B}. More precisely, it is the cosmic evolution of the source class in question that determines how much energy is fed into an electromagnetic cascade and eventually reappears in the GeV-band background radiation, relative to the UHECR energy flux at the ankle.

\subsection{The transition from galactic to extragalactic origin}
An open problem in cosmic-ray astrophysics is at what energy we observe
the transition from a Galactic to an extragalactic origin of particles.
The limit on inferred source power per unit baryon mass required to sustain
Galactic UHECR in the [4-40] EeV range that is imposed by the observed
anisotropy limits is smaller by nearly 3 orders of magnitude
than what is required for an extragalactic origin, as calculated in
\citep{egp10}, and
it corresponds to the power per unit mass of gamma rays from GRB \citep{ep11}.
This is not only confirmation of the hypothesis that UHECR beyond the ankle are
extragalactic, it also suggest that their sources are systems not persistently
present in the Galaxy. Any astrophysical source class, that is capable of accelerating
particles to very high energies and should exist in the Galaxy, may fall short of
accounting for the trans-ankle UHECR, but may nevertheless significantly contribute
to the observed cosmic-ray flux between the knee and the ankle. For example,
the numerical
coincidence fits the hypothesis of a GRB origin for the Galactic component of UHECR,
without invoking a much larger unseen energy reservoir for GRB.

The interesting question is the rate with which such sources appear in normal galaxies
such as the Milky Way. In other words, what is the role of intermittency?
Generally, GRBs in the Galaxy are expected every million years or so, the exact rate
depending on the beaming fraction and the detailed scaling of long GRB with star formation and metallicity. Therefore, only a small
number of GRB can contribute to the particle flux at the solar circle, and their relative
contribution depends on the location and explosion time of the GRB. Variations in the local
particle flux must be expected, and neither the particle spectrum from an individual GRB
nor the spectrum calculated for a homogeneous source distribution are good proxies.

\citet{pe11} have calculated the time-dependent transport of UHECR in the Galaxy,
assuming it can be described as isotropic diffusion. They find that
intermittency becomes serious if the mean free path for scattering exceeds
100~pc, unless
the source rate is much higher than 1 per Myr. On average, Galactic long GRB
need to contribute only
about $10^{37}$~erg/s in accelerated particles to fully account for the
observed particle
flux at $10^{18}$~eV, assuming a Bohmian mean free path at this energy.
UHECR from Galactic long GRB can meet the observational limits on anisotropy
only if the
mean free path for scattering is sufficiently small. Contributing the observed
sub-ankle particles
(at $10^{18}$~eV) requires Bohmian diffusion if the UHECR are as heavy as
carbon. A light composition
such as protons or helium requires sub-Bohmian diffusion, which is a highly unlikely situation for isotropic diffusion.

Much of the UHECR anisotropy arises from the expected location of long GRB in the inner Galaxy.
Observations of GRB host galaxies suggest that regions of low metallicity and active star formation
may be the preferred sites of long GRB \citep{2010AJ....140.1557L,2011arXiv1101.4418L}, which
may skew the galactocentric distribution of long GRB toward the outer Galaxy.
As there is no power problem with Galactic GRB, it may be worthwhile to also consider short GRB.
They provide supposedly less power as a population, but they may have a very extended
spatial distribution in the Galaxy \citep{2010ApJ...722.1946B},
leading to a reduced, but on account of intermittency not disappearing anisotropy.

These conclusions can be applied with little change to the case of an origin
of UHECRs
in SNRs, assuming very efficient magnetic-field amplification can increase their ability to accelerate particles to energies significantly higher than $1$~PeV \citep[e.g.][]{2010ApJ...718...31P}. The spatial distribution in the Galaxy of long GRB and SNR can be expected to be similar, and therefore the average anisotropy is the same for both long GRB and SNR.
If one combines such a galactic component with a
dip model, so that the galactic/extragalactic transition occurs below 1 EeV, Bohm diffusion and a mixed composition of the Galactic component may still be viable, given the systematic uncertainties in the measurements.
It would be highly desirable to improve anisotropy measurement between 0.1~EeV and
1~EeV, and likewise better constrain the composition.

\subsection{UHECR summary}

Recent progress in UHECR research has built on data from new large-scale observatories.
The interpretation of measurements of the composition, anisotropy, and spectrum of particles provides constraining links between these observables, that are further strengthened by new precision measurement of, e.g., extragalactic gamma-ray background emission. The very low anisotropy observed for EeV-scale particles provides a strong limit on the contribution of Galactic sources, if the composition is indeed light as
suggested by data. The anisotropy found above 60 EeV would be difficult to understand
if the particles were heavy, which is suggested by Auger data, but not HiRes. If these particles were light, dip models might be favorable which, however, must be carefully constructed to not overproduce the 50-GeV-scale gamma-ray background.

The main obstacle to further progress clearly is the systematic uncertainty arising from
the interpretation of the evolution of giant airshowers with
particle-physics models that are extrapolated over at least 1.5 decades in CoM energy
from the range teastable with manmade accelerator experiments.

\section{Particle acceleration at relativistic shocks}\label{rs}
Strong shocks occurring in astrophysical flows often generate power-law
distributions of very-high-energy particles. This is the origin of
most high-energy phenomena in astrophysics. The favored mechanism for the
generation of supra-thermal particles is the famous Fermi process. It
involves with the scattering of high-energy particles off magnetic
disturbances that allow them to cross the shock back and forth and
thus to gain energy. Many studies have been performed in the 80-ties and
90-ties by assuming pre-existing magnetic turbulence. However, it
turns out that the pre-existing turbulence is generally not strong
enough to account for the acceleration performance.  The nonthermal X-ray emission from
SNRs \citep{Cassam}, but see also \citet{2005ApJ...626L.101P}, revealed
that the magnetic field is strongly amplified in the vicinity of
the forward shock. Recent theoretical studies have shown that the penetration of
accelerated particles in the shock upstream flow can generate
magnetic turbulence that reaches a level much larger than the
intensity of the ambient mean field \citep{Bell04,Pel06}. In
producing turbulence the cosmic rays loose a fraction of its
global energy (about 10 percent of the incoming energy) but increases the
maximum energy of particles (cf. Section \ref{ssec:maxE}).
The turbulent field can reach an intensity of a few
hundreds of $\mu$G, much larger
than the value of a few $\mu$G of the ambient magnetic field in the Galaxy.

These results incited similar investigations for relativistic shocks. Very
encouraging results were obtained around the turn of the century
which extended the theory of Fermi process to the case of relativistic
shocks and predicted the formation of a power-law energy spectrum with
an index $s = 2.2-2.3$ and an acceleration time as fast as the Larmor
time \citep{BO98,Gal99,Kirk00,Ach01,Elli02,LP03}. But disappointment
came once the effect of the
ambient magnetic field had been taken into account, because it
inhibits the Fermi process even when one considers a strong Kolmogorov
turbulence \citep{Nie06, Lemo:06}.

In the following, it will be shown how the paradigm of the three
interdependent aspects of collisionless-shock physics successfully works
in the absence of any mean field: structure with a
partial reflection on a barrier, supra-thermal-particle generation,
magnetic-turbulence generation. Then the scattering issue in the
presence of a mean magnetic field will be addressed and the
requirement for circumventing the inhibition effect will be
stated. Then an unusual fact in astrophysics will be emphasized,
namely the necessity of considering some unavoidable micro-physics,
that turns out to be crucial not only for the relativistic shock
formation but also for making the Fermi process operative and
producing high energy particles.

\subsection{Successful Fermi process at very low magnetization}

The most favored process for the generation of supra-thermal power law
distributions is the Fermi process at shocks. Under astrophysical conditions
the plasma flow that experiences a shock is supposed to carry a frozen-in turbulent
magnetic field which allows particle scattering, and thereby permits particles
to gain energy at each Fermi cycle, i.e. a cycle
upstream-downstream-upstream or downstream-upstream-downstream.

At a non-relativistic shock of speed $\beta_s = V_s/c \ll 1$, the
average gain per cycle is small, $G = 1 + {4\over 3}{r-1\over
  r}\beta_s$ (where $r$ is the compression ratio, that reaches the
value 4 when the shock is adiabatic and strong). However this is
compensated by a large number of shock crossings; indeed the escape
probability (i.e. the probability for a particle to be entrained by
the downstream flow and to not come back to the shock front) is low,
$P_{\rm esc} = 4 \beta_s/r$; the return probability $P_{\rm ret}$ is thus
large. A power-law distribution of energy is set up with an index that
is a simple function of the compression ratio, in the non-relativistic
case:
\begin{equation}
s = 1- {\ln P_{\rm ret} \over \ln G} \simeq 1+{3 \over r-1} \ .
\end{equation}
Strong adiabatic shocks provide a particle spectrum with an universal
index, $s \simeq 2$, which is modified by losses, radiation losses for
the electrons, expansion or escape for protons. Subsequent to escape,
the spectrum is then steepened by
the effect of diffusive propagation and escape of particles from the
Galaxy.

A sizable fraction of the incoming energy flux is converted into cosmic ray pressure:
\begin{equation}
P_{\rm cr} = \xi_{\rm cr} \rho_u V_s^2 \, \, {\rm with} \, \, \xi_{\rm
  cr} \sim 0.1 \ .
\end{equation}

The successive Fermi cycles produce a precursor of supra-thermal
particles (mostly protons) of large extension (the diffusion length
increases with the particle energy) and this penetration in the
upstream medium (the ambient medium for an external shock) triggers
MHD turbulence through two types of streaming instability, one is
resonant and has been considered for many years
\citep[see for instance][]{1982A&A...116..191M}, the other is
non-resonant and has been
considered more recently \citep{Bell04,Pel06}, as briefly discussed
earlier. That
latter case is quite interesting, first because it is a simple and
robust mechanism based on the supplementary Lorentz force associated
with the plasma current that compensates the cosmic-ray current,
second because it leads to a turbulent field of large intensity;
indeed this latter can become much larger than the ambient
magnetic field. The theory indicates that the fraction of incoming
energy flux converted into magnetic energy can reach $\xi_B \sim
\beta_s$, which is a few percent in SNRs, where one
defines
\begin{equation}
{B_{\rm rms}^2 \over 4\pi} = \xi_B \rho_u V_s^2 \ .
\end{equation}
A very important remark is that the efficiency of the Fermi process
depends on the efficiency of the scattering process. By the way, the
mechanism of Fermi acceleration is a simple process, but the
scattering, that controls the efficiency of the acceleration process, is the main issue.

\begin{table}[htdp]
\caption{Comparison non-relativistic shocks and relativistic shocks.}
\begin{center}
\begin{tabular}{|c|c|}
\hline {\bf At non-relativistic shocks} & {\bf At relativistic shocks} \\
\hline weak escape probability & significant escape probability \\
\hline many cycles of weak energy gain & few cycles of large energy gain \\
\hline power law distribution $\epsilon^{-s}$ with $s \simeq 2$ & power law distribution $\epsilon^{-s}$ with $s \sim 2.3$ \\
\hline upstream distribution weakly anisotropic & upstream distribution strongly anisotropic \\
\hline partial reflection at shock front & partial reflection at shock front \\
\hline generation of MHD turbulence upstream  & generation of e.m. micro-turbulence upstream \\
\hline acceleration time $t_{acc} \sim \tau_s/\beta_s^2$ & acceleration time $t_{acc} \sim \tau_s$ \\
\hline
\end{tabular}
\end{center}
\label{default}
\end{table}%

As for relativistic shocks, there are similarities and some
differences with the non-relativistic ones, as summarized in Table
1. There are strong arguments that there is a significant generation
of magnetic turbulence at the external shock of a GRB \citep{Liwax06}
and there is an obvious power-law distribution of
ultra-relativistic electrons that synchrotron radiate, with an index
compatible with the theory of the Fermi process at ultra-relativistic
shocks ($s=2.2-2.3$). The ambient magnetic field is very low and at
first approximation can be neglected. A remarkable work was published
by \citet{2008ApJ...682L...5S} that fully validates the paradigm,
combining three fundamental processes: the formation of a collisionless
relativistic shock front with reflected particles, the generation of
magnetic turbulence and the generation of a power-law distribution
through the Fermi process. This is a PIC (Particles In Cell) simulation
of the development of a collisionless shock in a pair plasma
(electrons and positrons) that runs with a Lorentz factor $\Gamma_s$
of a few tens ($\Gamma_s \equiv (1-\beta_s^2)^{-1/2}$). The flow of
reflected particles interacts with the flow of passing particles
leading to streaming-type instabilities, and the Weibel branch of
instability describes the formation of intense small-scale magnetic
filaments. The relevant scale of the physics is the inertial length
(or skin depth) $\delta \equiv {c \over \omega_p}$. The spatial growth
of the magnetic micro-turbulence produces a partial reflection of the
incoming particles, which allows the formation of a shock front, and
self-consistently, the reflected particles generate the required level
of micro-turbulence.  Similarly as the non-relativistic case,
conversion parameters $\xi_{\rm cr}, \xi_B$ can be defined in the
ultra-relativistic case:
\begin{eqnarray}
P_{\rm cr} & =  & \xi_{\rm cr} \rho_u \Gamma_s^2 c^2 \\
{B_{\rm rms}^2 \over 4\pi} & = & \xi_B \rho_u \Gamma_s^2 c^2
\end{eqnarray}
And the simulations indicate that $\xi_{\rm cr} \sim 0.1$ and $\xi_B \sim
1-10 \%$, similarly to the non-relativistic case. The supra-thermal spectrum obtained
in the simulation is close to the theoretical prediction with an index $s \simeq 2.4$.
Similar results were obtained later with PIC simulation involving a
plasma of electrons and ions of $(10\dots 100)\, m_e$ \citep{Siro:09}.

\subsection{Opening phase space with finite magnetization}

Many astrophysical shocks form in a plasma having a significant
magnetization.  The physics becomes more complex with a finite
ambient mean field; it is controlled by the important
``magnetization'' parameter $\sigma$:
\begin{equation}
\sigma \equiv {B_{t,f}^2 \over 4\pi \rho_u \Gamma_s^2 c^2} = {B_0^2 \sin^2 \theta_B \over 4\pi \rho_u c^2} \ ,
\end{equation}
where $B_0$ is the field measured in the upstream flow frame
(generally the ambient field), and $B_{t,f}$ is the transverse
component of the mean field measured in the shock frame. Like in
non-relativistic shocks, the angle of the field lines with respect to
the shock normal is very important. But whereas most non-relativistic
shocks are in the so-called ``sub-luminal'' configuration, i.e. that
the angle $\theta_B$ is not too close to $90^0$ and thus particles can
flow along the field lines, in ultra-relativistic shocks,
it suffices that the field angle $\theta_B$ be larger than $1/
\Gamma_s$ to prevent the return of particles to the upstream region. A
generic ultra-relativistic shock is thus ``supra-luminal'', and the
magnetic field in the front frame can be considered as almost
perpendicular, because its transverse component is amplified by a
factor $\Gamma_s$. This field orientation is a serious hindrance for
the development of Fermi cycles. {Neglecting for the time being
any scattering process in a putative turbulence superimposed on the
background field, the particle kinematics can be described as
follows. } A particle that enters the downstream flow of speed $c/3$
is dragged by the frozen in magnetic field and cannot easily come back
upstream; it can be shown that it can come back just one time
\citep{Lemo:06}. Once upstream, it eventually comes back downstream,
but in a subset of phase space that does not allow it to make a second
cycle. {Now, it} might be thought that a strong turbulence could
provide efficient scattering allowing it to make several
cycles. {However, the typical interstellar turbulent field} with
a large-scale coherence length behaves like an ordered magnetic field
for such particles, because their penetration length upstream ($\ell_p
= m_pc^2/\Gamma_s eB_0$, measured in the co-moving upstream frame) is
much shorter than the coherence length of turbulence \citep{Lemo:06}.
{In self-generated small scale turbulence, scattering might be
efficient enough to trigger Fermi acceleration, see below.}

The coherence length $\ell_c$ is formally defined as the range of the
field correlation using the self-correlation function, $C(r)$.  For an
isotropic turbulent state we can write (it can easily be properly
modified in the case of anisotropic turbulence):
\begin{equation}
\ell_c \equiv \int_0^{\infty} C(r)dr \ ;
\end{equation}
which can be expressed as an integral over the turbulence
spectrum, and one finds that for a spectrum proportional to $k^{-\beta}$, the
correlation length corresponds to large wavelengths for $1<
\beta < 2$, as is the case of a Kolmogorov spectrum; for $ 0 \leq
\beta \leq 1$, the coherence length is in the shortest-wavelengths
part of the spectrum.

Moreover, the expected duration of the cycle would be much shorter than
the eddy turn-over time of large-scale vortices. The requirements for efficient
scattering off magnetic turbulence are quite challenging
\citep{2009MNRAS.393..587P}, for not only the intensity of the turbulent
field must be much larger than the mean field, but also the coherence
length must be shorter than a Larmor radius. When a scattering process
develops, phase space is opened for operating a Fermi
process if the scattering frequency is larger than the
Larmor pulsation in the mean field. Short-scale turbulence leads to a
scattering frequency $\nu_s \propto \epsilon^2$, whereas the Larmor
pulsation $\omega_L \propto \epsilon$; thus the range of particle
energies for which the phase space is unlocked and Fermi process
operative, is such that $\epsilon < \epsilon_{scatt} \equiv Ze({\bar
  B^2/B_0})\ell_c$.

At high magnetization (say $\sigma > 0.03$) the shock is formed by
generation of an intense coherent wave through a Synchrotron Maser
Instability due to a resonance with the loop of reflected particles
\citep{1991PhFlB...3..818H,Gall92}.  The electromagnetic wave
propagating downstream is damped by synchrotron resonance and produces
a thermal distribution. The wave propagating upstream carries away a
fraction $\sim 0.1 \sigma$ of the incoming energy in the case of an
$e^+-e^-$-plasma; in a $p^+-e^-$-plasma, an electrostatic wake field
is generated that heats the electrons up to equipartition while
slowing down protons
\citep{1991PhFlB...3..818H,2011ApJ...726...75S}. The formation of a
power-law distribution \citep{2008ApJ...672..940H} has not been
confirmed as far as we know. {The final word has not been given
on these issues of course, because the simulations have been
conducted so far in 1D  \citep{1991PhFlB...3..818H,Gall92} or 2D
\citep{2011ApJ...726...75S} over a limited amount of time. What
happens in a more realistic 3D simulation, the dimensionality of
which should allow more efficient cross-field transport, or on
longer timescales, remain to be seen \citep[see][]{1998ApJ...509..238J}.}

According to \citet{2011ApJ...726...75S}, at lower magnetization,
nothing happens except the thermalization of protons ($T_p \simeq 0.2
\, \Gamma_s m_pc^2$), until the magnetization reaches a very low
critical value at which the Fermi process starts. {As the
magnetization decreases, indeed the precursor length scale
increases, to the point where plasma microinstabilities triggered by
the suprathermal particle population self-generate a small scale
turbulence that can sustain the Fermi process (Lemoine \& Pelletier
2010).}  Actually, one needs a very low magnetic field to obtain an
upstream penetration length of supra-thermal particles large enough
for having a significant interaction of those particles with the
incoming plasma and having a growth of micro-instabilities. The Fermi
process works with the magnetic component of micro-turbulence at the
inertial scale $\sim \delta \equiv c/ \omega_{pi}$. In principle it
starts at even smaller scale, the inertial scale of electrons, however
electrons are efficiently heated by the electric component of
micro-turbulence and then the precursor becomes composed of electrons
and protons of similar relativistic mass, like a pair plasma. This is
a very interesting outcome that simplifies the physics which rapidly
evolves towards conditions similar to those occurring in a pair
plasma. Thus the PIC simulations of pair plasma are also valuable to
understand the physics of shocks in electron-proton plasmas. Then a
distribution function displaying a thermal part and a supra-thermal
part with a power law is obtained.

The transition towards the Fermi process is determined by the
micro-instabilities that can grow when the upstream penetration of
reflected particles is long enough. The fastest instabilities
\citep[Buneman
instability, Oblique Two-Stream instability, see][]{2004PhRvE..70d6401B} seem
to essentially pre-heat the incoming electrons almost up to
equipartition with protons. However, more simulations are necessary to
clarify this important point. The generation of magnetic micro-turbulence by
the Weibel instability, which is also studied in laboratory experiments,
is thought to be the main ingredient to form collisionless shocks
and to produce the Fermi process; however this is also under study
by PIC simulations.  The generation of magnetic micro-turbulence
occurs when the magnetization parameter falls below the following
critical value \citep{2011MNRAS.417.1148L}, as confirmed by
numerical simulations \citep{2011ApJ...726...75S}:
\begin{equation}
\sigma < \sigma_{crit} \equiv {\xi_{cr} \over \Gamma_s^2} \ .
\end{equation}
Numerical simulations show that the level reached by that Weibel
turbulence is such that $\xi_B = 1-10 \%$, which insures shock
formation and Fermi process. Then there exists a large energy range
for particle scattering when $\sigma \ll \xi_B^2$.

\subsection{The micro-physics aspect of GRB termination shocks}\label{sec:acc}

The main issue with Fermi processes based on the scattering off
micro-turbulence is that the scattering frequency decreases as
$E^{-2}$.  The performance of Fermi processes at non-relativistic
shocks is determined by the scattering off large-scale, say Kolmogorov,
turbulence which is fairly slow (much slower than the Larmor pulsation in the
mean field) but decreases only as $E^{-1/3}$. Thus, if we
compare the Fermi process at relativistic shocks with the process at
non-relativistic shock, this is like the hare and the tortoise: the
scattering, and thus the acceleration rhythm, at relativistic shocks
is very fast at low energy and decreases rapidly as energy increases,
whereas, at non-relativistic shocks, it is slow at low energy but
continues at higher energies with a moderate decline of its
efficiency.

\subsubsection{Electron acceleration and radiation}

The external shock that drives the afterglow emission of GRBs
{may give} rise to an efficient acceleration of electrons
{if the external medium is weakly magnetized, for the reasons
discussed previously}. If {electrons} thermalize with protons
(as reasonably expected), their temperature is already very high at
the beginning of the afterglow: $T_e \sim T_p \simeq 0.2 \Gamma_s
m_pc^2$, which corresponds to a few tens of GeV. Intense short-scale
magnetic turbulence develops because the interstellar magnetization
parameter is very low, $\sigma \sim 10^{-9}$, whereas the critical
value $\sigma_{\rm crit} \sim 10^{-6}$, with $\Gamma_s \sim 300$.

What kind of radiation can be expected in such small-scale field, much
more intense than the mean field? This depends on a so-called
``wiggler" parameter $a$:
\begin{equation}
a \equiv \frac{e\,B_{\rm rms}\, \ell_c}{m_e\,c^2} \sim \xi_B^{1/2}\,
\Gamma_s\, {m_p \over m_e} \ .
\end{equation}
This parameter measures the capability of the magnetic force to
deflect a relativistic electron of Lorentz factor $\gamma$ by an angle
$1/\gamma$ (which is the reason why $\gamma$ does not appear in
the definition). If $a >1$, then the magnetic field produces a single
deflection of the electron in the emission cone of half angle $1/
\gamma$, whereas if $a<1$ the electron can undergo several wiggles
in the emission cone. When $a$ is large, the emission behaves like
normal synchrotron radiation in a mean field, except that there is no
polarization. When $a$ is small, the emission is of ``jitter" type
\citep{2000ApJ...540..704M}. Thus the emission caused by shocked and accelerated
electrons at a relativistic shock is ``synchrotron-like", and the
analysis of the emitted spectrum provides a diagnostic of the
magnetic turbulence.

It is quite remarkable that there exists an almost universal energy
limit for the electron radiating in the intense small scale field
\citep[in
agreement with][]{2010ApJ...710L..16K}:
\begin{equation}
\gamma_{\rm max} \approx (\frac{4\pi\, e^2\, \ell_c}{\sigma_T\, m_e\, c^2})^{1/3}
\simeq (\frac{m_p}{n\,m_e\,r_e^3})^{1/6} \approx 10^6 \ .
\end{equation}
The corresponding maximum energy for the  photons emitted inthe
quasi-homogeneous field is
\begin{equation}\label{emax}
E_{\gamma, \rm max} \sim \sqrt{\pi}\, \xi_B^{1/2}\, {\Gamma_s^2 \over
  \gamma_{\rm max}}\, {{m_p\, c^2} \over \alpha_f} \sim
2 \times ({\xi_B \over 10^{-2}})^{1/2}\, \left({\Gamma_s \over 300}\right)^2\quad
 {\rm GeV} \ ,
\end{equation}
where $\alpha_f$ is the fine structure constant. The account for
magnetic fluctuations of scales larger than the synchrotron emission
formation length results in the photon spectra extended beyond the
limit given by Eq.(\ref{emax}) \citep[see][]{2012MNRAS.421L..67B}.
Thus a single synchrotron-like spectrum extending up to several GeV,
even possibly a few tens, can be expected and is in fact compatible
with observations. So the performance of relativistic shocks for
electron acceleration and radiation {appears very
satisfactory}. The conversion factor into radiation is $\xi_{\rm
rad} \sim \xi_B\, \sigma_T\, n_0 \,r_s\, <\gamma_e^2>$, and at the
beginning of the afterglow $\xi_{\rm rad} \sim \xi_B \sim 1-10 \%$.

\subsubsection{Proton acceleration limited by the fast decay of scattering}
Protons are expected to be accelerated at least as efficiently
as electrons at ultra-relativistic shock waves. However, as
mentioned previously, the {ultra-relativistic} Fermi process
{appears unable to push protons up to energies in excess of $\sim
10^{17}-10^{18}\,$eV}, because the scattering time and thus the
acceleration time increase with $E^2$ in the self-generated turbulent
field, or scale with $E$, but then in the background unamplified
field.

Using one or the other, one does not find numbers significantly
different from the limit associated to the mean field {discussed
in Sec.~\ref{ssec:maxE}} (Eq.~\ref{emax1}): $E_{\rm max} = Z\, \Gamma_S\,
e\,B_0\, R_S \simeq Z \times (0.3 \cdot 10^7\ {\rm GeV})$.  Thus,
although an energy of order $10^{16}$~eV is achieved, which is
something, the result is far from reaching the UHE-range.

Precise performances of mildly or sub-relativistic shocks are not yet
known and require more numerical simulations. However, some reasonable
estimates are permitted by extrapolating what we know about the two
extremes: non-relativistic and ultra-relativistic shocks. The main
guess is that we can expect a magnetic-field amplification at shocks
with a conversion factor $\xi_B = 1-10 \%$, occurring in MHD regime
without severe limitation due to the super-luminal configuration,
especially for oblique internal shocks (termination shocks in the hot
spots of FR2 jets might be super-luminal). These assumptions can be
applied to internal shocks of AGN jets (in particular in Blazars
jets), and to internal shocks of GRBs, as already discussed in
Sec.~\ref{sec:UHECR}.

\subsection{Conclusion and Prospect}
\label{sec:shock1}

The triangular dependence of collisionless shock structure with a
reflecting barrier for a part of incoming particles, with generation
of supra-thermal particles and the generation of magnetic turbulence
is a successful paradigm that applies to astrophysical shocks, both
non-relativistic and relativistic. Numerical and theoretical works are
making significant progress for both understanding the physics and
providing quantitative results useful for astrophysical
investigations. This includes not only the spectral index and cut off
of the distribution of accelerated particles, but also the efficiency
factors for the conversion into cosmic rays, magnetic turbulence and
radiation. We have seen only the
beginning of this line of study, which requires more PIC simulations
and new types of hybrid codes involving relativistic MHD coupled with
PIC codes for cosmic rays.

The new results that have already been obtained are important. First,
the strong amplification of the magnetic field at SNRs received
theoretical and numerical support; the astrophysical consequences are
interesting, especially for our understanding of the Galactic
contribution of the cosmic-ray spectrum. Secondly, {current
state-of-the-art PIC simulations indicate that the Fermi process
does not operate at ultra-relativistic shocks with magnetization of
order unity}, which is supposed to be a frequent situation in
high-energy astrophysics, as for instance in FR2 hot spots, in
blazars, in pulsar wind nebulae.  {Such simulations need however
to be extended both to higher dimensionality and to larger
space-time domains before a definite conclusion can be
reached. {In particular, the issue of the non stationarity and/or corrugation 
of the shock front
in relativistic regime should be investigated. Also the role of magnetic 
reconnections in the shock vicinity
is a very important new topic whose investigation is just starting}. 
Whether and how acceleration proceeds in the
mildly-relativistic regime also remains open for study.}  Thirdly
the radiation processes that operate in most high-energy astrophysical
sources involve relativistic electrons scattered in an intense
short-scale magnetic turbulence; this leads to a renewed interest in
the radiation physics with a view to use it as a diagnostic of the
magnetic turbulent state.

The new trend in these topics is the important role imputed to
micro-physics phenomena, which have a direct astrophysical impact.
These developments incite interest in several other communities,
including space-plasma physics, laser-plasma physics, astroparticle,
and high-energy astrophysics. We live in exciting times.

\begin{acknowledgements}
A.M.B. was supported in part by the Russian government grant
11.G34.31.0001 to Sankt-Petersburg State Politechnical University,
and also by the RAS and RAS Presidium Programs and by the RFBR grant
11-02-12082. N.G. acknowledges scientific discussions and assistance
by J. Cannizzo.  M.P. is supported by the `Helmholtz Alliance for
Astroparticle Phyics HAP´ funded by the Initiative and Networking
Fund of the Helmholtz Association. M.L. and G.P. acknowledge
financial support of the PEPS/PTI program of the CNRS-INP.
\end{acknowledgements}
\bibliographystyle{svjour}

\begin{thebibliography}{aps-nameyear}

\bibitem[Abbasi et al.(2010)]{2010PhRvL.104p1101A} Abbasi, R.~U.,
Abu-Zayyad, T., Al-Seady, M., et al.\ 2010, Physical Review Letters, 104,
161101

\bibitem[Abdo et al.(2008)]{s8} Abdo, A.A., et al. 2010, Phys. Rev. Lett., 104, 101101

\bibitem[Abdo et al.(2009a)]{2009Sci...323.1688A} Abdo, A.~A., Ackermann,
M., Arimoto, M., et al.\ 2009a, Science, 323, 1688

\bibitem[Abdo et al.(2009b)]{2009Natur.462..331A} Abdo, A.~A., Ackermann,
M., Ajello, M., et al.\ 2009b, \nat, 462, 331

\bibitem[Abraham et al.(2010b)]{2010PhRvL.104i1101A} Abraham, J., Abreu, P.,
Aglietta, M., et al.\ 2010b, Physical Review Letters, 104, 091101

\bibitem[Abraham et al.(2010a)]{2010PhLB..685..239A} Abraham, J., Abreu, P.,
Aglietta, M., et al.\ 2010a, Physics Letters B, 685, 239

\bibitem[Abraham et al.(2008)]{2008APh....29..188P}
Abraham, J., Abreu, P., et al.\ 2008,
Astroparticle Physics, 29, 188

\bibitem[Abraham et al.(2004)]{2004NIMPA.523...50A} Abraham, J., Aglietta,
M., Aguirre, I.~C., et al.\ 2004, Nucl. Instr. and Methods in Phys.
Research A, 523, 50

\bibitem[Abreu et al.(2011)]{2011APh....34..627P}
Abreu, P., Aglietta, M., et al.\ 2011,
Astroparticle Physics, 34, 627

\bibitem[Abreu et al.(2010)]{2010APh....34..314T}
Abreu, P., Aglietta, M., et al.\ 2010,
Astroparticle Physics, 34, 314

\bibitem[Acciari et al.(2009)]{Acci:09} Acciari, V. A., Aliu, E., Arlen, T., et al. 2009, Science, 325, 444

\bibitem[Acciari et al.(2011)]{Acci:11} Acciari, V.~A., Aliu, 
E., Arlen, T., et al.\ 2011, \apj, 738, 25 

\bibitem[Achterberg(1983)]{A83} Achterberg, A. \ 1983, AA, 119, 274

\bibitem[Achterberg et al.(2001)]{Ach01} Achterberg, A.,
Gallant, Y.~A., Kirk, J.~G., \& Guthmann, A.~W.\ 2001, MNRAS, 328,
393

\bibitem[Aharonian et al.(1999)]{Ahar:99}
Aharonian, F.\ A., Akhperjanian, A.\ G., Barrio, J.\ A., et al. 1999, A\&A, 349,  11

\bibitem[Aloisio et al.(2011)]{Aloisio11} Aloisio, R., 
Berezinsky, V., \& Gazizov, A.\ 2011, Astroparticle Physics, 34, 620 

\bibitem[Aloisio et al.(2007)]{s6} Aloisio, R.,
Berezinsky, V., Blasi, P., et al.\ 2007, Astroparticle Physics, 27, 76

\bibitem[Apel et al.(2009)]{2009APh....31...86A} Apel, W.~D., Arteaga,
J.~C., Badea, A.~F., et al.\ 2009, Astroparticle Physics, 31, 86

\bibitem[Arons(2003)]{s3} Arons, J.\ 2003, \apj, 589, 871

\bibitem[Barthelmy et al.(2005)]{2005Natur.438..994B} Barthelmy, S.~D.,
Chincarini, G., Burrows, D.~N., et al.\ 2005, \nat, 438, 994

\bibitem[Bednarz
\& Ostrowski(1998)]{BO98} Bednarz, J., \& Ostrowski, M.\ 1998, Physical Review Letters, 80, 3911

\bibitem[Bell(2005)]{Bell05} Bell, A.~R.\ 2005, \mnras, 358,
181

\bibitem[Bell(2004)]{Bell04} Bell, A.~R.\ 2004, \mnras, 353,
550

\bibitem[Bell \& Lucek(2001)]{bl01} Bell, A.R., Lucek, S.G.\ 2001, MNRAS 321, 433

\bibitem[Berezhko(2008)]{2008ApJ...684L..69B} Berezhko, E.~G.\ 2008, \apjl,
684, L69

\bibitem[Berezinsky et al.(2006)]{2006PhRvD..74d3005B} Berezinsky, V.,
Gazizov, A., \& Grigorieva, S.\ 2006, \prd, 74, 043005

\bibitem[Berger(2010)]{2010ApJ...722.1946B} Berger, E.\ 2010, \apj, 722,
1946

\bibitem[Berger et al.(2005)]{2005Natur.438..988B} Berger, E., Price,
P.~A., Cenko, S.~B., et al.\ 2005, \nat, 438, 988

\bibitem[Blandford
\& Znajek(1977)]{1977MNRAS.179..433B} Blandford, R.~D., \& Znajek, R.~L.\ 1977, \mnras, 179, 433

\bibitem[Biermann \& Strittmatter(1987)]{s1}
Biermann, P.L., \& Strittmatter, P.A. 1987, ApJ, 322, 643

\bibitem[Blandford
\& McKee(1976)]{1976PhFl...19.1130B} Blandford, R.~D., \& McKee, C.~F.\ 1976, Physics of Fluids, 19, 1130

\bibitem[Bloom et al.(2006)]{2006ApJ...638..354B} Bloom, J.~S., Prochaska,
J.~X., Pooley, D., et al.\ 2006, \apj, 638, 354

\bibitem[Bret et al.(2004)]{2004PhRvE..70d6401B} Bret, A., Firpo, M.-C.,
\& Deutsch, C.\ 2004, \pre, 70, 046401

\bibitem[Burrows et al.(2006)]{2006ApJ...653..468B} Burrows, D.~N., Grupe,
D., Capalbi, M., et al.\ 2006, \apj, 653, 468

\bibitem[{{Bykov}(2001)}]{bykov01}
{Bykov}, A.M. 2001, Space Science Reviews 99, 317

\bibitem[{{Bykov} and {Meszaros}(1996)}]{bm96}
{Bykov}, A.M., {Meszaros}, P. 1996, \apjl 461, L37,

\bibitem[{{Bykov} et~al.(2012){Bykov}, {Pavlov}, {Artemyev}, and
  {Uvarov}}]{2012MNRAS.421L..67B}
{Bykov}, A.M., {Pavlov}, G.G., {Artemyev}, A.V., {Uvarov}, Y.A. 2012, MNRAS Letters 421, L67

\bibitem[{{Bykov} and {Treumann}(2011)}]{bt11}
{Bykov}, A.M., {Treumann}, R.A. 2011, Astron. Astroph. Reviews 19, 42

\bibitem[{{Bykov} and {Toptygin}(1993)}]{bt93}
{Bykov}, A.M., {Toptygin}, I.N. 1993, Physics-Uspekhi 36, 1020

\bibitem[Bykov et al.(2011)]{bykov05} Bykov, A.M., Osipov, S.M. Ellison, D.C. \ 2011, MNRAS, 410,
39
\bibitem[Cassam-Chena{\"i} et
al.(2004)]{Cassam} Cassam-Chena{\"i}, G., Decourchelle, A., Ballet, J., et al.\ 2004, \aap, 414, 545


\bibitem[Celotti et al.(2001)]{Celo:01} Celotti, A.,
Ghisellini, G., \& Chiaberge, M.\ 2001, MNRAS, 321, L1

\bibitem[Costa et al.(1997)]{1997Natur.387..783C} Costa, E., Frontera, F.,
Heise, J., et al.\ 1997, \nat, 387, 783

\bibitem[{{Daigne} and {Mochkovitch}(1998)}]{dm98}
{Daigne}, F., {Mochkovitch}, R. 1998, \mnras 296, 275

\bibitem[Eichler(1981)]{ei81} Eichler, D.\ 1981, ApJ 244, 711

\bibitem[Eichler
\& Pohl(2011)]{ep11} Eichler, D., \& Pohl, M.\ 2011, \apjl, 738, L21

\bibitem[Eichler et al.(2010)]{egp10} Eichler, D., Guetta, D., \& Pohl, M.\ 2010, ApJ 722, 543

\bibitem[Ellison
\& Double(2002)]{Elli02} Ellison, D.~C., \& Double, G.~P.\ 2002, Astroparticle Physics, 18, 213

\bibitem[{{Ferrand} and {Marcowith}(2010)}]{fm10}
{Ferrand}, G., {Marcowith}, A., 2010, \aap 510, A101

\bibitem[Fong et al.(2010)]{2010ApJ...708....9F} Fong, W., Berger, E.,
\& Fox, D.~B.\ 2010, \apj, 708, 9

\bibitem[Forman and Drury(1983)]{fordr83} Forman, M.A. \& Drury, L. O'C.\ 1983, International Cosmic Ray Conference, 2, 267F

\bibitem[Fox et al.(2005)]{2005Natur.437..845F} Fox, D.~B., Frail, D.~A.,
Price, P.~A., et al.\ 2005, \nat, 437, 845

\bibitem[Frail et al.(1997)]{1997Natur.389..261F} Frail, D.~A., Kulkarni,
S.~R., Nicastro, L., Feroci, M., \& Taylor, G.~B.\ 1997, \nat, 389, 261

\bibitem[Frail et al.(2001)]{2001ApJ...562L..55F} Frail, D.~A., Kulkarni,
S.~R., Sari, R., et al.\ 2001, \apjl, 562, L55


\bibitem[Gallant
\& Achterberg(1999)]{Gal99} Gallant, Y.~A., \& Achterberg, A.\ 1999,
\mnras, 305, L6


\bibitem[Gallant et al.(1992)]{Gall92} Gallant, Y.~A.,
Hoshino, M., Langdon, A.~B., Arons, J., \& Max, C.~E.\ 1992, \apj, 391, 73

\bibitem[Garson et al.(2010)]{Gars:10} Garson, A.\ B., III, Baring, M.\ G., Krawczynski, H.\ 2010, ApJ, 722, 358

\bibitem[Gehrels et al.(2004)]{2004ApJ...611.1005G} Gehrels, N.,
Chincarini, G., Giommi, P., et al.\ 2004, \apj, 611, 1005

\bibitem[Gehrels et al.(2005)]{2005Natur.437..851G} Gehrels, N., Sarazin,
C.~L., O'Brien, P.~T., et al.\ 2005, \nat, 437, 851

\bibitem[Gehrels et
al.(2009)]{2009ARA&A..47..567G} Gehrels, N., Ramirez-Ruiz, E., \& Fox, D.~B.\ 2009, \araa, 47, 567

\bibitem[{{Giannios} and {Spruit}(2005)}]{gs05}
{Giannios}, D., {Spruit}, H.C. 2005, \aap 430, 1

\bibitem[Giannios et al.(2010)]{Gian:10}
Giannios, D., Uzdensky, D.\ A., Begelman, M.\ C.\ 2010, MNRAS, 402, 1649

\bibitem[{{G{\"o}tz} et~al.(2009){G{\"o}tz}, {Laurent}, {Lebrun}, {Daigne}
  et~al.}]{goetzea09}
{G{\"o}tz}, D., {Laurent}, P., {Lebrun}, F., {Daigne}, F., et~al. 2009, \apjl 695, L208



\bibitem[Grupe et al.(2006)]{2006ApJ...653..462G} Grupe, D., Burrows,
D.~N., Patel, S.~K., et al.\ 2006, \apj, 653, 462

\bibitem[Harris \& Krawczynski(2002)]{Harr:02} Harris, D.~E., \& Krawczynski, H.\ 2002, ApJ, 565, 244

\bibitem[Harris \& Krawczynski(2006)]{Harr:06} Harris, D.~E., \& Krawczynski, H.\ 2006, ARAA, 44, 463

\bibitem[Hillas(1984)]{hillas}  Hillas, M.\ 1984, Ann. Reviews of Astr. and Ap., 22, 425

\bibitem[Hjorth et al.(2005)]{2005Natur.437..859H} Hjorth, J., Watson, D.,
Fynbo, J.~P.~U., et al.\ 2005, \nat, 437, 859

\bibitem[Hoshino(2008)]{2008ApJ...672..940H} Hoshino, M.\ 2008, \apj, 672,
940

\bibitem[Hoshino
\& Arons(1991)]{1991PhFlB...3..818H} Hoshino, M., \& Arons, J.\ 1991, Physics of Fluids B, 3, 818

\bibitem[Jones et al.(1998)]{1998ApJ...509..238J} Jones, F.~C., Jokipii, 
J.~R., \& Baring, M.~G.\ 1998, \apj, 509, 238 

\bibitem[Kachelrie{\ss} et al.(2010)]{Kach:10} Kachelrie{\ss}, M., Ostapchenko, S.,
Tomàs, R.\ 2010, Publications of the Astronomical Society of Australia, 27, 482

\bibitem[Kachelrie{\ss} \& Semikoz(2006)]{s7}
Kachelriess, M., \& Semikoz, D. V. 2006, Phys. Lett. B, 634, 143

\bibitem[Kaneko et al.(2007)]{2007ApJ...654..385K} Kaneko, Y.,
Ramirez-Ruiz, E., Granot, J., et al.\ 2007, \apj, 654, 385

\bibitem[Kaneko et al.(2008)]{2008ApJ...677.1168K} Kaneko, Y.,
Gonz{\'a}lez, M.~M., Preece, R.~D., Dingus, B.~L.,
\& Briggs, M.~S.\ 2008, \apj, 677, 1168

\bibitem[Kang et al.(1997)]{s2}
Kang, H., Rachen, J. P., \& Biermann, P. L. 1997, MNRAS, 286, 257

\bibitem[Kawai et al.(2008)]{2008NuPhS.175..221K} Kawai, H., Yoshida, S.,
Yoshii, H., et al.\ 2008, Nuclear Physics B Proceedings Supplements, 175,
221

\bibitem[{{Kirk} et~al.(2009){Kirk}, {Lyubarsky}, and {Petri}}]{klp09}
{Kirk} JG, {Lyubarsky} Y, {Petri} J (2009)  In: Astrophysics and
Space Science Library (ed. {W~Becker}), vol.
  357 of \emph{Astrophysics and Space Science Library}, pp. 421



\bibitem[Kirk
\& Reville(2010)]{2010ApJ...710L..16K} Kirk, J.~G., \& Reville, B.\ 2010, \apjl, 710, L16

\bibitem[Kirk et al.(2000)]{Kirk00} Kirk, J.~G., Guthmann,
A.~W., Gallant, Y.~A., \& Achterberg, A.\ 2000, \apj, 542, 235

\bibitem[Kirk et al.(1998)]{Kirk:98} Kirk, J.~G., Rieger, F.~M.,
\& Mastichiadis, A.\ 1998, A\&A, 333, 452

\bibitem[Klebesadel et al.(1973)]{1973ApJ...182L..85K} Klebesadel, R.~W.,
Strong, I.~B., \& Olson, R.~A.\ 1973, \apjl, 182, L85

\bibitem[Kotera
\& Olinto(2011)]{2011ARA&A..49..119K} Kotera, K., \& Olinto, A.~V.\ 2011, \araa, 49, 119

\bibitem[Kouveliotou et al.(1993)]{1993ApJ...413L.101K} Kouveliotou, C.,
Meegan, C.~A., Fishman, G.~J., et al.\ 1993, \apjl, 413, L101

\bibitem[Krawczynski et al.(2002)]{Kraw:02}
Krawczynski, H., Coppi, P.\ S., Aharonian, F.\ 2002, MNRAS, 336, 721

\bibitem[Krawczynski et al.(2001)]{Kraw:01}
Krawczynski, H., Sambruna, R., Kohnle, A., et al. 2001, ApJ, 559, 187

\bibitem[Kuiper et
al.(2001)]{2001A&A...378..918K} Kuiper, L., Hermsen, W., Cusumano, G., et al.\ 2001, \aap, 378, 918

\bibitem[Kumar
\& Barniol Duran(2010)]{2010MNRAS.409..226K} Kumar, P., \& Barniol Duran, R.\ 2010, \mnras, 409, 226

\bibitem[{{Kumar}(1999)}]{kumar99}
{Kumar}, P. 1999, \apjl 523, L113

\bibitem[Lagage \& Cesarsky(1983a)]{lc1} Lagage, P.O., Cesarsky, C.\ 1983a, A\&A 125, 249

\bibitem[Lagage \& Cesarsky(1983b)]{lc2} Lagage, P.O., Cesarsky, C.\ 1983b, A\&A 118, 223

\bibitem[Le \& Dermer(2007)]{ld07} Le, T., Dermer, C.D.\ 2007, ApJ 661, 394

\bibitem[Lemoine \& Pelletier(2011b)]{LP11b} Lemoine, M. \& Pelletier,
  G.\ 2011b, arXiv:1111.7110

\bibitem[Lemoine
\& Pelletier(2011a)]{2011MNRAS.417.1148L} Lemoine, M., \& Pelletier, G.\ 2011a, \mnras, 417, 1148

\bibitem[Lemoine \& Pelletier(2010)]{Lemo:10}Lemoine, M., Pelletier, G.\ 2010, MNRAS, 402, 321

\bibitem[Lemoine
\& Waxman(2009)]{2009JCAP...11..009L} Lemoine, M., \& Waxman, E.\ 2009, \jcap, 11, 9

\bibitem[Lemoine et al.(2006)]{Lemo:06} Lemoine, M., Pelletier,
G., \& Revenu, B.\ 2006, \apjl, 645, L129

\bibitem[Lemoine
\& Pelletier(2003)]{LP03} Lemoine, M., \& Pelletier, G.\ 2003, \apjl, 589, L73

\bibitem[Levesque(2011)]{2011arXiv1101.4418L} Levesque, E.~M.\ 2011, 
American Institute of Physics Conference Series, 1358, 271 

\bibitem[Levesque et al.(2010)]{2010AJ....140.1557L} Levesque, E.~M.,
Kewley, L.~J., Berger, E., \& Jabran Zahid, H.\ 2010, \aj, 140, 1557

\bibitem[Li
\& Waxman(2006)]{Liwax06} Li, Z., \& Waxman, E.\ 2006, \apj, 651, 328

\bibitem[Lucek \& Bell(2000)]{lb00} Lucek, S.G., Bell, A.R.\ 2000, MNRAS 314, 65

\bibitem[{{Lyutikov} and {Blandford}(2004)}]{lb04}
{Lyutikov}, M., {Blandford}, R. 2004,  In:
  Astronomical Society of the Pacific Conference Series (ed. {M~Feroci,
  F~Frontera, N~Masetti, \& L~Piro}), vol. 312 of \emph{Astronomical Society of
  the Pacific Conference Series}, p. 449



\bibitem[MacFadyen
\& Woosley(1999)]{1999ApJ...524..262M} MacFadyen, A.~I., \& Woosley, S.~E.\ 1999, \apj, 524, 262

\bibitem[Malkov \& Diamond(2001)]{md01} Malkov, M.A., Diamond, P.H.\ 2001, Phys. Plasmas 8 (5), 2401

\bibitem[McConnell et al.(2002)]{2002ApJ...572..984M} McConnell, M.~L.,
Zdziarski, A.~A., Bennett, K., et al.\ 2002, \apj, 572, 984

\bibitem[McKenzie
\& Voelk(1982)]{1982A&A...116..191M} McKenzie, J.~F., \& Voelk, H.~J.\ 1982, \aap, 116, 191

\bibitem[McKinney(2005)]{2005ApJ...630L...5M} McKinney, J.~C.\ 2005, \apjl,
630, L5

\bibitem[McKinney
\& Narayan(2007a)]{2007MNRAS.375..513M} McKinney, J.~C., \& Narayan, R.\ 2007a, \mnras, 375, 513

\bibitem[McKinney
\& Narayan(2007b)]{2007MNRAS.375..531M} McKinney, J.~C., \& Narayan, R.\ 2007b, \mnras, 375, 531

\bibitem[{{McKinney} and {Uzdensky}(2012)}]{mu12}
{McKinney} JC, {Uzdensky} DA 2012,  \mnras 419, 573

\bibitem[Medvedev(2000)]{2000ApJ...540..704M} Medvedev, M.~V.\ 2000, \apj,
540, 704

\bibitem[Medvedev
\& Loeb(1999)]{1999ApJ...526..697M} Medvedev, M.~V., \& Loeb, A.\ 1999, \apj, 526, 697

\bibitem[{{M{\'e}sz{\'a}ros}(2006)}]{mesz06}
{M{\'e}sz{\'a}ros} P (2006)  Reports on Progress in Physics
  69, 2259


\bibitem[{{Meszaros} and {Rees}(1997)}]{mr97}
{Meszaros} P, {Rees} MJ (1997)  \apjl 482, L29

\bibitem[{{Mimica} and {Aloy}(2012)}]{ma12}
{Mimica} P, {Aloy} MA (2012) MNRAS :2427

\bibitem[Mizuno et~al.(2011)]{Mizunoea10}
Mizuno, Y., Pohl, M., 
Niemiec, J., et al.\ 2011, \apj, 726, 62 


\bibitem[{{Murase} et~al.(2012){Murase}, {Asano}, {Terasawa}, and
  {M{\'e}sz{\'a}ros}}]{muraseea12}
{Murase} K, {Asano} K, {Terasawa} T, {M{\'e}sz{\'a}ros} P (2012)
 \apj 746:164

\bibitem[Murase et al.(2009)]{2009PhRvD..79j3001M} Murase, K.,
M{\'e}sz{\'a}ros, P., \& Zhang, B.\ 2009, \prd, 79, 103001

\bibitem[Murase et al.(2008)]{s5} Murase, K., Inoue, S.,
\& Nagataki, S.\ 2008, \apjl, 689, L105

\bibitem[Murase et al.(2006)]{2006ApJ...651L...5M} Murase, K., Ioka, K.,
Nagataki, S., \& Nakamura, T.\ 2006, \apjl, 651, L5

\bibitem[Niemiec et al.(2006)]{Nie06} Niemiec, J., Ostrowski,
M., \& Pohl, M.\ 2006, \apj, 650, 1020


\bibitem[{{Nishikawa} et~al.(2010){Nishikawa}, {Nimiec}, {Medvedev}, {Zhang}
  et~al.}]{Nishikawaea10}
{Nishikawa} KI, {Nimiec} J, {Medvedev} M, {Zhang} B, et~al. (2010)
 International
  Journal of Modern Physics D 19:715--721

\bibitem[Norman et al.(1995)]{Nea95} Norman, C., Melrose, D. \&
  Achterberg, A. \ 1995, ApJ, 454, 60

\bibitem[{{Panaitescu} et~al.(1999){Panaitescu}, {Spada}, and
  {M{\'e}sz{\'a}ros}}]{psm99}
{Panaitescu} A, {Spada} M, {M{\'e}sz{\'a}ros} P (1999)  \apjl
522:L105

\bibitem[Pe'Er et al.(2009)]{2009PhRvD..80l3018P} Pe'Er, A., Murase, K.,
\& M{\'e}sz{\'a}ros, P.\ 2009, \prd, 80, 123018

\bibitem[Pelletier et al.(2009)]{2009MNRAS.393..587P} Pelletier, G.,
Lemoine, M., \& Marcowith, A.\ 2009, \mnras, 393, 587

\bibitem[Pelletier et
al.(2006)]{Pel06} Pelletier, G., Lemoine, M., \& Marcowith, A.\ 2006, \aap, 453, 181

\bibitem[Perlman et al.(1999)]{Perl:99} Perlman, E.\ S., Biretta, J.\ A., Zhou, F.,
Sparks, W.\ B., and Macchetto, F.\ D.\ 1999, Astron.\ J., 117, 2185

\bibitem[{{Piran}(2004)}]{piran04}
{Piran} T (2004)  Reviews of Modern Physics
  76:1143

\bibitem[Plotnikov et al.(2011)]{Plotnikov11} Plotnikov, I.,
  Pelletier, G., \& Lemoine, M. \ 2011, A\&A, 532, 68

\bibitem[Pohl
\& Eichler(2011)]{pe11} Pohl, M., \& Eichler, D.\ 2011, \apj, 742, 114 

\bibitem[Pohl et al.(2005)]{2005ApJ...626L.101P} Pohl, M., Yan, H., 
\& Lazarian, A.\ 2005, \apjl, 626, L101 

\bibitem[Pohl et al.(2002)]{pls02} Pohl, M., Lerche, I., Schlickeiser, R.\ 2002, A\&A, 383, 309

\bibitem[Pohl \& Schlickeiser(2000)]{ps00} Pohl, M., Schlickeiser, R.\ 2000, A\&A, 354, 395

\bibitem[Ptuskin et al.(2011)]{Ptuskin11} Ptuskin, V. S., Rogovaya,
  S. I., \& Zirakashvili, V. N. \ 2011, arXiv:1105.4491

\bibitem[Ptuskin et al.(2010)]{2010ApJ...718...31P} Ptuskin, V.,
Zirakashvili, V., \& Seo, E.-S.\ 2010, \apj, 718, 31

\bibitem[Rachen
\& Biermann(1993)]{1993A&A...272..161R} Rachen, J.~P., \& Biermann, P.~L.\ 1993, \aap, 272, 161

\bibitem[{{Rees} and {Meszaros}(1994)}]{rm94}
{Rees} MJ, {Meszaros} P (1994)  \apjl 430:L93--L96

\bibitem[Reville et al.(2006)]{2006PPCF...48.1741R} Reville, B., Kirk,
J.~G., \& Duffy, P.\ 2006, Plasma Physics and Controlled Fusion, 48, 1741

\bibitem[Salvaterra et al.(2009)]{2009Natur.461.1258S} Salvaterra, R.,
Della Valle, M., Campana, S., et al.\ 2009, \nat, 461, 1258

\bibitem[Sari et al.(1998)]{1998ApJ...497L..17S} Sari, R., Piran, T.,
\& Narayan, R.\ 1998, \apjl, 497, L17

\bibitem[Sironi
\& Spitkovsky(2011)]{2011ApJ...726...75S} Sironi, L., \& Spitkovsky, A.\ 2011, \apj, 726, 75

\bibitem[Sironi \& Spitkovsky(2009)]{Siro:09}
Sironi, L., Spitkovsky, A.\ 2009, ApJ, 698, 1523

\bibitem[Skilling(1975)]{S75} Skilling, J. \ 1975, MNRAS, 173, 245

\bibitem[Spitkovsky(2008)]{2008ApJ...682L...5S} Spitkovsky, A.\ 2008,
\apjl, 682, L5

\bibitem[{{Spruit} et~al.(2001){Spruit}, {Daigne}, and {Drenkhahn}}]{sdd01}
{Spruit} HC, {Daigne} F, {Drenkhahn} G (2001)  \aap 369:694--705

\bibitem[Takahara(1990)]{1990PThPh..83.1071T} Takahara, F.\ 1990, Progress
of Theoretical Physics, 83, 1071

\bibitem[Takahashi et al.(2000)]{Taka:00} Takahashi, T., et al.\ 2000, ApJL, 542, 105

\bibitem[Tanvir et al.(2009)]{2009Natur.461.1254T} Tanvir, N.~R., Fox,
D.~B., Levan, A.~J., et al.\ 2009, \nat, 461, 1254

\bibitem[Tavecchio et al.(2000)]{Tave:00} Tavecchio, F.,
Maraschi, L., Sambruna, R.~M., \& Urry, C.~M.\ 2000, \apjl, 544, L23

\bibitem[{{Thompson}(1994)}]{thompson94}
{Thompson} C (1994)  \mnras 270:480

\bibitem[{{Usov}(1992)}]{usov92}
{Usov} VV (1992)  \nat 357:472--474


\bibitem[Turner et al.(2002)]{turner} Turner, M.S., et al. 2002,
{\it Report to the National Academy of Science}

\bibitem[van Paradijs et al.(1997)]{1997Natur.386..686V} van Paradijs, J.,
Groot, P.~J., Galama, T., et al.\ 1997, \nat, 386, 686

\bibitem[Vietri(1995)]{1995ApJ...453..883V} Vietri, M.\ 1995, \apj, 453,
883

\bibitem[Villasenor et al.(2005)]{2005Natur.437..855V} Villasenor, J.~S.,
Lamb, D.~Q., Ricker, G.~R., et al.\ 2005, \nat, 437, 855

\bibitem[Vladimirov et al.(2006)]{veb06} Vladimirov A., Ellison D.C., Bykov, A.\ 2006, ApJ 652, 1246

\bibitem[Wang et al.(2007)]{2007PhRvD..76h3009W} Wang, X.-Y., Razzaque, S.,
M{\'e}sz{\'a}ros, P., \& Dai, Z.-G.\ 2007, \prd, 76, 083009

\bibitem[Waxman(2010)]{2010arXiv1010.5007W} Waxman, E.\ 2010,
arXiv:1010.5007

\bibitem[Waxman(2005)]{W05} Waxman, E. \ 2005, Physica Scripta, T121, 147

\bibitem[Waxman(2004)]{wax2004} Waxman, E.\ 2004, ApJ 606, 988

\bibitem[Waxman(1995)]{s4} Waxman, E.\ 1995, Physical
Review Letters, 75, 386

\bibitem[Wentzel(1974)]{W74} Wentzel, D. G. \ 1974, ARAA, 12, 71

\bibitem[{{Yamada} et~al.(2010){Yamada}, {Kulsrud}, and {Ji}}]{yk10}
{Yamada}, M., {Kulsrud}, R., {Ji}, H. 2010, Reviews of
  Modern Physics 82, 603

\bibitem[{{Yonetoku} et~al.(2011){Yonetoku}, {Murakami}, {Gunji}, {Mihara}
  et~al.}]{Yonetokuea11}
{Yonetoku}, D., {Murakami}, T., {Gunji}, S., {Mihara}, T., et~al. 2011,
 \apjl 743,  L30,

\bibitem[Zhang et al.(2003)]{2003ApJ...586..356Z} Zhang, W., Woosley,
S.~E., \& MacFadyen, A.~I.\ 2003, \apj, 586, 356

\bibitem[{{Zhang} and {Yan}(2011)}]{zy11}
{Zhang}, B., {Yan}, H. 2011,  \apj 726, 90


\end{thebibliography}




\end{document}